\documentclass[
 reprint,
 amsmath,amssymb,
 aps,
]{revtex4-2}
\usepackage{graphicx}
\usepackage{dcolumn}
\usepackage{bm}
\usepackage{graphicx}
\usepackage{amsmath}  %ydj
\usepackage{hyphenat} 

\usepackage[utf8]{inputenc}
\usepackage{graphicx}% Include figure files
\usepackage{dcolumn}% Align table columns on decimal point
\usepackage{bm}% bold math
\usepackage{float}
\usepackage{amsmath}
\usepackage{graphicx,subfigure,epsfig}
\usepackage{hyperref}
\hypersetup{
    colorlinks=true,
    citecolor=blue,
    linkcolor=blue,
    filecolor=magenta,
    urlcolor=cyan,}

\begin{document}

\preprint{APS/123-QED}

\title{Observational properties and quasinormal Modes of the Hayward black Hole surrounded by a cloud of
strings}

\author{Qi-Qi Liang$^{1}$}
\author{Ziqiang Cai$^{1}$}
\author{Dong Liu$^{2}$}
\author{Zheng-Wen Long$^{1}$}
\email[Corresponding author: ]{zwlong@gzu.edu.cn}

\affiliation{$^{1}$College of Physics, Guizhou University, Guiyang, 550025, China}
\affiliation{$^{2}$Department of Physics, Guizhou Minzu University, Guiyang, 550025, China}

\begin{abstract}

In this work, we explored the Hayward black hole surrounded by a cloud of strings, with a focus on the effects of the regularization parameter $l$ and the string cloud parameter $a$ on its observational properties and quasinormal modes (QNMs). Utilizing the spacetime metric and geodesic equations, we calculated several geometric quantities characterizing the black hole. 
To visualize the observational appearance of the accretion disk, we employed the Novikov--Thorne model to simulate both its primary and secondary images.
%of the black hole's accretion disk.
Furthermore, we analyzed the QNMs of the black hole under scalar and electromagnetic perturbations for different parameter values.
The results indicate that as the regularization parameter $l$ increases, the outer horizon radius $r_{+}$, photon-sphere radius $r_{\text{ph}}$, critical impact parameter $b_{c}$, and innermost stable circular orbit $r_{\text{isco}}$ exhibit a gradual decrease, 
while the inner horizon radius $r_{-}$ and the real part of the QNMs frequency $\omega_{r}$ increase.
In contrast, as the string cloud parameter $a$ increases, $r_{+}$, $r_{\text{ph}}$, $b_{c}$, and $r_{\text{
isco}}$ demonstrate a rapid increase, whereas $r_{-}$ and $\omega_{r}$ decrease. 
In both cases, the absolute value of the imaginary part of the QNMs frequency decreases with the increase $l$ or $a$.
This work offers a theoretical foundation for understanding the coupling between regular black holes and surrounding string clouds. 

\end{abstract}

\maketitle

%\tableofcontents
\section{Introduction}

General relativity, as the fundamental theoretical framework for strong gravitational fields, has successfully explained classical observational puzzles such as the perihelion precession of Mercury and the gravitational redshift of the Sun~\cite{Einstein:1915bz,Will:2018mcj,Delva:2018ilu,RocaCortes:2014}, and has further predicted extreme physical phenomena such as gravitational waves and black holes~\cite{Einstein:1918btx,Schwarzschild:1916uq}. In 2015, the Laser Interferometer Gravitational-Wave Observatory (LIGO) collaboration announced the detection of gravitational waves~\cite{LIGOScientific:2016aoc}, and in 2019, the Event Horizon Telescope (EHT) released the first image of a black hole (M87*)~\cite{EventHorizonTelescope:2019ths}. These observations are in remarkable agreement with theoretical simulations based on general relativity, thereby confirming these predictions one by one. Traditional black holes generally suffer from singularity problems, where the curvature diverges at $r=0$ and the laws of physics break down~\cite{Penrose:1964wq,Piazza:2025uxm}. This is in clear conflict with the requirement of finiteness imposed by quantum gravity at the smallest scales. To overcome this difficulty, theoretical physicists have proposed the concept of regular black holes, which theoretically eliminate the spacetime singularity of conventional black holes and provide models whose spacetime structure remains physically reasonable in all regions~\cite{Ansoldi:2008jw,Capozziello:2024ucm}. The first such model, the Bardeen black hole~\cite{Bardeen:1968wq}, was constructed based on nonlinear electromagnetic fields. Subsequently, other regular solutions were proposed, such as the Ay\'on-Beato--Garc\' ia black hole~\cite{Ayon-Beato:2000mjt} and the Hayward black hole~\cite{Hayward:2005gi}, all of which approach a ``de Sitter'' spacetime in the central region, thereby keeping the curvature finite.

Black holes in the universe rarely exist in a ``vacuum-isolated'' and stationary state. They are usually surrounded by plasma, dark matter, or exotic matter~\cite{Perivolaropoulos:2021jda}. These matter fields couple to the black hole spacetime through the stress–energy tensor and can significantly modify the observable characteristics of black holes~\cite{Belkhadria:2025lev,NunesdosSantos:2025alw,Jusufi:2025slp,Jha:2025xjf}. For example, when a Hayward black hole is surrounded by quintessence, the corresponding metric solution reveals how variations in the core parameters influence the black hole properties, including the analysis of photon trajectories via null geodesics, the Hawking temperature and greybody factors, as well as holographic descriptions of the black hole constructed through the AdS/CFT correspondence. These approaches highlight, from different perspectives, the modifications induced by quintessence on the physical properties of the Hayward black hole~\cite{Yashwanth:2024suw,Pedraza:2020uuy,Li:2024owp,Al-Badawi:2023lke,He:2024bll}. The Hayward black hole has also been generalized to the rotating case, where its physical properties have been further investigated~\cite{Amir:2015pja,Abdujabbarov:2016hnw,Jusufi:2018jof,Neves:2014aba,Bambi:2013ufa,Tsukamoto:2014tja}.
String clouds, as a macroscopic configuration in string theory composed of a large number of low-temperature strings, can mimic the complex quantum-gravity environment surrounding black holes. When a string cloud couples to the black hole spacetime, it introduces an additional stress–energy source, thereby modifying the metric solution, horizon structure, and dynamical behavior of the black hole~\cite{Mustafa:2022xod,Yang:2023agi,Li:2020zxi,Herscovich:2010vr,Al-Badawi:2024cby,Gogoi:2022ove}. 
In particular, it has been shown that when Bardeen and Hayward black holes are surrounded by a string cloud, their original regularity may be destroyed, leading to the emergence of new singularities or discontinuities~\cite{Rodrigues:2022zph,Vishvakarma:2023csw,Nascimento:2023tgw}. This phenomenon not only deepens our understanding of the applicability of regular black hole models but also provides an important testing ground for exploring possible quantum-gravity corrections.

In the study of black hole properties, the Observational properties and QNMs serve as crucial observational probes for revealing the nature of spacetime and testing gravitational theories. The shadow is a dark region formed by the combined effects of light capture and gravitational lensing in a strong gravitational field, and in realistic astrophysical scenarios, it is often surrounded by a bright accretion disk~\cite{Narayan:2019imo,Falcke:1999pj}. Therefore, black hole shadows and accretion disks can directly reflect the spacetime structure of black holes. Quasinormal modes, are the damped oscillations generated when a black hole is perturbed by external fields. These oscillations gradually decay over time, and their characteristic parameters - such as frequency and damping rate - are determined solely by the intrinsic properties of the black hole, including its mass and spin, much like a ``fingerprint'' of the black hole. Owing to these features~\cite{Kokkotas:1999bd,Berti:2009kk}, both the shadow and QNMs have become key tools for testing fundamental physical theories and deepening our understanding of black holes.

The QNMs of Hayward black holes have been studied in different backgrounds.\cite{Guo:2021bhr,Lin:2013ofa,Malik:2024tuf,Konoplya:2023ppx,Konoplya:2022hll,Konoplya:2024hfg}.However, studies on the observational properties and quasinormal mode characteristics of the Hayward black hole surrounded by a string cloud remain absent. This issue is not only of theoretical significance but may also provide novel approaches for identifying the distribution of external matter around black holes in future observations.
The structure of this paper is arranged as follows. In Sec~\ref{sec:II}, we review the spacetime of the Hayward black hole surrounded by a string cloud, calculate the geometric parameters of the black hole spacetime for different values of $a$ and $l$ through geodesic analysis, and simulate the accretion disk images based on the Novikov--Thorne model. In  Sec~\ref{sec:III}, we investigate the related properties of the model under scalar and electromagnetic field perturbations. Finally, Sec~\ref{sec:IV} presents a short  conclusion.

\begin{figure*}[t]
  \centering
  \includegraphics[width=0.45\textwidth]{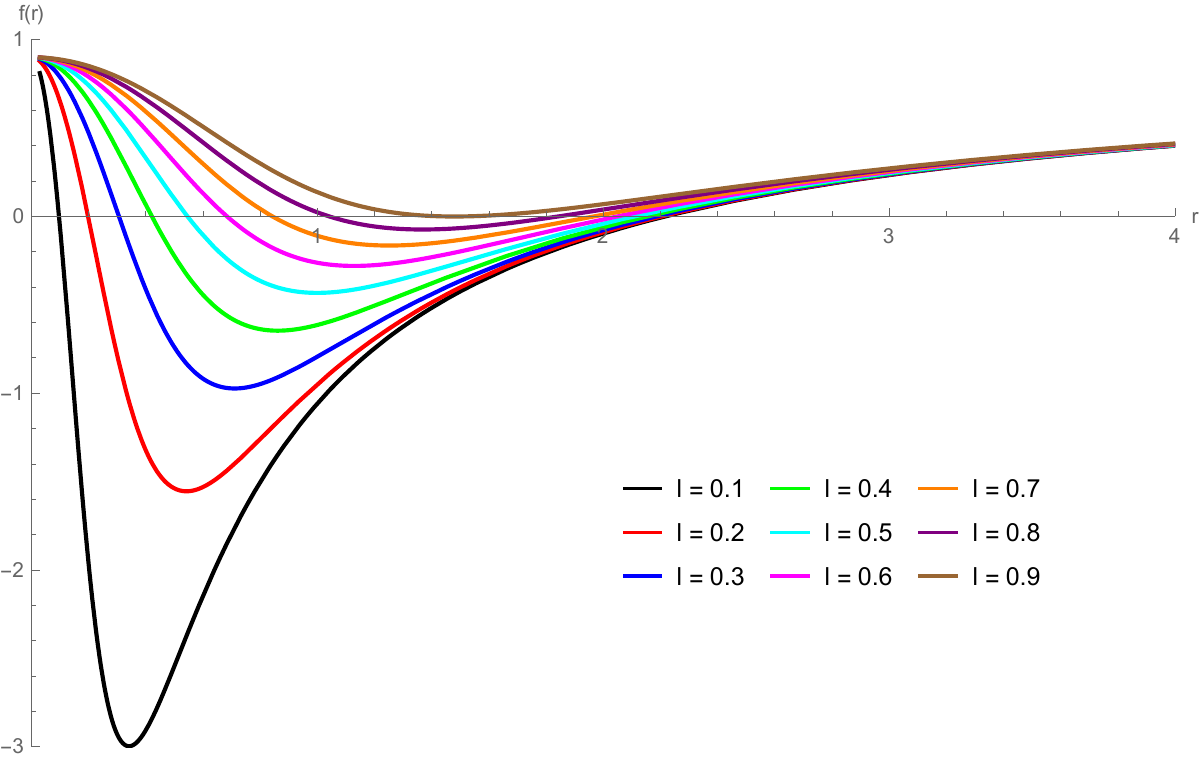}\hfill
  \includegraphics[width=0.45\textwidth]{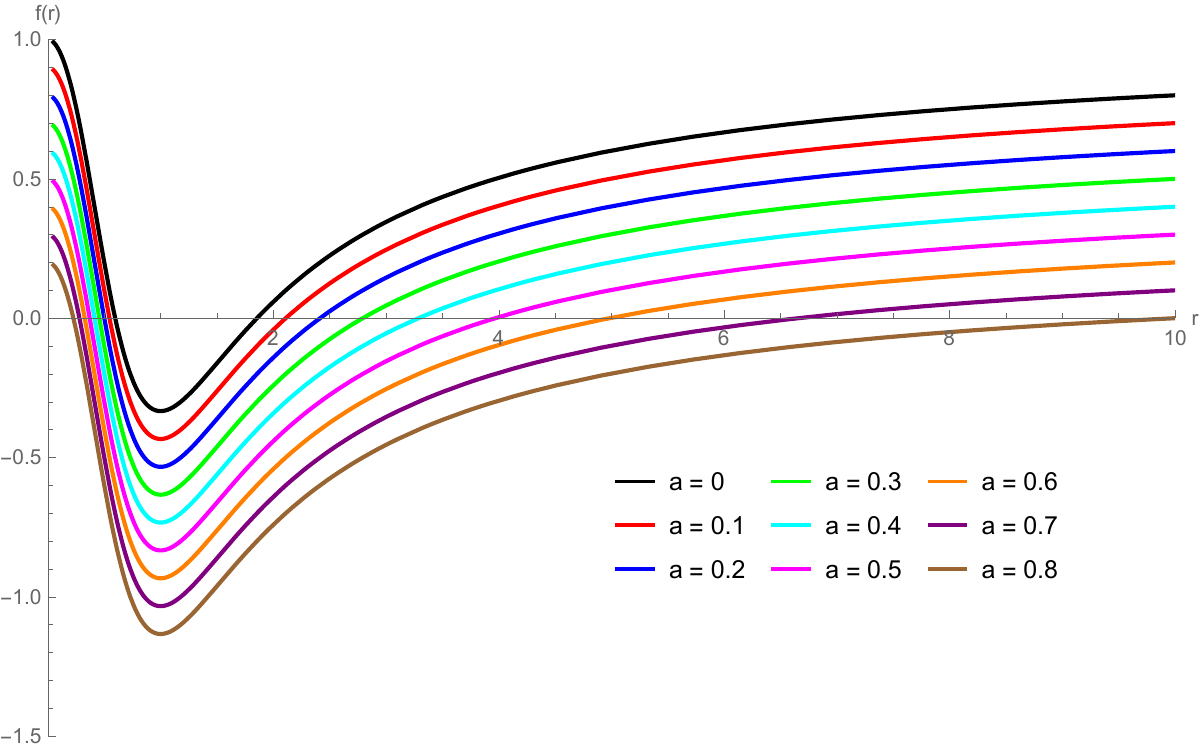}
  \caption{Variation curves of $f(r)$ with $r$ under different parameters, with $a = 0.1$ (left) and $l = 0.5$ (right)}
  \label{fig:1}
\end{figure*}
\section{Observational properties}
\label{sec:II}
\subsection{Geodesic Equation}

For a Hayward black hole surrounded by a string cloud, its geometric shape is given by the following line element~\cite{Nascimento:2023tgw}:

\begin{equation}
    ds^{2}=-f(r)dt^{2}+\frac{1}{f(r)}dr^{2}+r^{2}(d\theta^{2}+\sin^{2}\theta d\varphi^{2})
\label{equ:1}
\end{equation}

\begin{equation}
f(r) = 1 - a - \frac{2mr^2}{r^3 + 2l^2m}
 \label{equ:2}   
\end{equation}
where $l$ is the regularization parameter in the Hayward metric, $a$ is the string cloud parameter with a value range of  $0 < a < 1$, and $m$ represents the mass of the Hayward black hole.  When $a = 0$, the string cloud disappears and the black hole reduces to the Hayward black hole; when $a = 0$ and $l = 0$, the black hole returns to the Schwarzschild black hole.

This section adopts the geometric unit system where \(G = M = c = 1\). The metric function \( f(r) \) determines the position of the event horizon. To intuitively explore the relationship between \( f(r) \) and the parameters \( a \) and \( l \), we plot Fig. \ref{fig:1}. As shown in the figure, with the increase of the parameter $l$, the inner horizon of the black hole increases significantly while the outer horizon decreases slowly. In contrast, as the parameter $a$ increases, the inner horizon decreases slowly whereas the outer horizon increases significantly. 
The specific $r$ values of the inner and outer horizons are listed in Table \ref{tab:1}.

To gain a clearer understanding of the spatiotemporal properties of a black hole, we analyze the black hole's shadow through geodesics. We also consider the case of the equatorial plane for geodesic analysis. The Lagrangian of a particle in a static spherically symmetric spacetime is expressed as:

\begin{equation}
\mathcal{L} = \frac{1}{2} g_{\mu\nu}\frac{dx^\mu}{d\lambda} \frac{dx^\nu}{d\lambda}=\left( -f(r) \dot{t}^2 + \frac{\dot{r}^2}{f(r)} + r^2 \dot{\theta}^2 + r^2 \sin^2 \theta \, \dot{\phi}^2 \right)
\label{equ:3}
\end{equation}
where $\lambda$ is the affine parameter, the symbol ``$\cdot$'' denotes the derivative with respect to the affine parameter. For photons, $\mathcal{L} = 0$. The particle has two conserved quantities: energy momentum \( E \) and angular momentum \( L \). These two quantities can be used to define the impact parameter \( b \), with the expression \( b = \frac{L}{E} \). The impact parameter describes the initial aiming distance between the particle and the center of the black hole, and determines the trajectory of the particle in the gravitational field of the black hole.

Considering a spherically symmetric black hole and repar\-ametrizing the affine parameter by letting \( \lambda' = L \lambda \), the orbital equation for a photon approaching the black hole can be expressed as the motion equation in the orbital plane:

\begin{equation}
\dot{t} = \dfrac{1}{b f(r)} 
\label{equ:4}
\end{equation}
\begin{equation}
\dot{\phi} = \pm \dfrac{1}{r^2} 
\label{equ:5}
\end{equation}
\begin{equation}
\dot{r}^2 = \dfrac{1}{b^2} - \dfrac{f(r)}{r^2}
\label{equ:6}
\end{equation}
where $\frac{f(r)}{r^2}$ is defined as the effective of the photon. When \( r \) satisfies the following condition, the corresponding position is the radius of the photon sphere.

\begin{equation}
\left. V_{\text{eff}} \right|_{r = r_{\text{ph}}} = 0
\label{equ:7}
\end{equation}

\begin{equation}
\left. \frac{dV_{\text{eff}}}{dr} \right|_{r = r_{\text{ph}}} = 0
\label{equ:8}
\end{equation}
From the above two equations, we can obtain the corresponding impact parameter

\begin{equation}
b_{\rm c} = \frac{r_{\text{ph}}}{\sqrt{f(r_{\text{ph}})}}
\label{equ:9}
\end{equation}
The motion of photons approaching the black hole can be determined by the value of \( b_{\rm c} \): when \( b < b_{\rm c} \), the light ray will fall into the black hole; when \( b = b_{\rm c} \), the light ray will orbit the black hole in a circular path; when \( b > b_{\rm c} \), the light ray will be bound and eventually escape to infinity.

To study the bending of light near a black hole. By introducing the variable \( u = 1/r \), using equations~\ref{equ:5} and~\ref{equ:6}, we can derive that the trajectory \( u(\phi) \) of a photon on the equatorial plane satisfies:
\begin{equation}
G(u) := \left( \frac{du}{d\phi} \right)^2 = \frac{1}{b^2} - u^2 f\left( \frac{1}{u} \right)
\label{equ:10}
\end{equation}

In black hole spacetimes, besides the study of photon motion, the motion of timelike particles is also of great importance. The matter in accretion disks is composed of timelike particles, and their orbits determine the geometric boundaries of the accretion disk. In particular, the innermost Stable Circular Orbit defines the inner boundary of a thin accretion disk. For the geodesics of timelike particles with $\mathcal{L} = -1/2$, a discussion similar to that for photons leads to the radial motion equation of timelike particles as follows:
\begin{equation}
\dot{r}^2 = \frac{1}{b^2} - \frac{f(r)}{r^2} - \frac{f(r)}{L^2}
\label{equ:11}
\end{equation}

The radius \( r_{\text{isco}} \) of the innermost stable circular orbit for timelike particles satisfies the following equation:

\begin{equation}
\left. U_{\rm eff} \right|_{r = r_{\rm isco}} = \left. \frac{dU_{\rm eff}}{dr} \right|_{r = r_{\rm isco}} = \left. \frac{d^2 U_{\rm eff}}{dr^2} \right|_{r = r_{\rm isco}} = 0
\label{equ:12}
\end{equation}

\begin{table*}[htbp]
\centering
\caption{The values of parameters related to the black hole spacetime geometry under different parameters.}
\label{tab:1}       % Give a unique label
\begin{tabular}{llllllllll}
\hline
     & \multicolumn{9}{c}{$a$ = 0.1}       \\
\hline
\multicolumn{1}{c}{ } & \multicolumn{1}{c}{$l$ = 0.1} & \multicolumn{1}{c}{$l$ = 0.2} & \multicolumn{1}{c}{$l$ = 0.3} & \multicolumn{1}{c}{$l$ = 0.4} & \multicolumn{1}{c}{$l$ = 0.5} & \multicolumn{1}{c}{$l$ = 0.6}  & \multicolumn{1}{c}{$l$ = 0.7} & \multicolumn{1}{c}{$l$ = 0.8} & \multicolumn{1}{c}{$l$ = 0.9}\\
\hline

$r_{-}$   & 0.097  & 0.1988 & 0.3065 & 0.4216 & 0.5462 & 0.6842 & 0.8429 & 1.041  & 1.4299 \\
$r_{+}$   & 2.2182 & 2.2058 & 2.1845 & 2.1532 & 2.1099 & 2.0511 & 1.9696 & 1.847  & 1.5319 \\
$r_{\text{ph}}$ & 3.3297 & 3.3188 & 3.3002 & 3.2733 & 3.2372 & 3.1903 & 3.1301 & 3.052 & 2.949  \\
$b_{c}$   & 6.0825 & 6.0726 & 6.0557 & 6.0315 & 5.9994 & 5.9583 & 5.9066 & 5.8420 & 5.7603 \\
$r_{isco}$   & 6.6617 & 6.6468 & 6.6215 & 6.5856 & 6.5382 & 6.4783 & 6.4046 & 6.3149 & 6.2063 \\
\hline
     & \multicolumn{9}{c}{$l$ = 0.5}       \\
\hline
\multicolumn{1}{c}{} & \multicolumn{1}{c}{$a$ = 0} & \multicolumn{1}{c}{$a$ = 0.1} & \multicolumn{1}{c}{$a$ = 0.2} & \multicolumn{1}{c}{$a$ = 0.3} & \multicolumn{1}{c}{$a$ = 0.4} & \multicolumn{1}{c}{$a$ = 0.5}  & \multicolumn{1}{c}{$a$ = 0.6} & \multicolumn{1}{c}{$a$ = 0.7} & \multicolumn{1}{c}{$a$ = 0.8} \\
\hline
$r_{-}$   & 0.597  & 0.5462 & 0.5000 & 0.4564 & 0.4138 & 0.3712 & 0.3271 & 0.2798 & 0.2262 \\
$r_{+}$   & 1.8546 & 2.1099 & 2.4142 & 2.7930 & 3.2871 & 3.9682 & 4.9798 & 6.6554 & 9.9950 \\
$r_{\text{ph}}$ & 2.878  & 3.23721 & 3.6756 & 4.2296 & 4.9593 & 5.9719 & 7.48213 & 9.9899 & 14.9956 \\
$b_{c}$   & 5.0928 & 5.9994 & 7.1905 & 8.8146 & 11.135 & 14.6626 & 20.5152 & 31.6069 & 58.0861 \\
$r_{isco}$   & 5.839 & 6.5382 & 7.3996 & 8.4953 & 9.9444 & 11.9616 & 14.9755 & 19.9862 & 29.9939 \\
\hline
\end{tabular}
\end{table*}

\begin{figure*}[t]
\centering
  \includegraphics[width=0.5\textwidth]{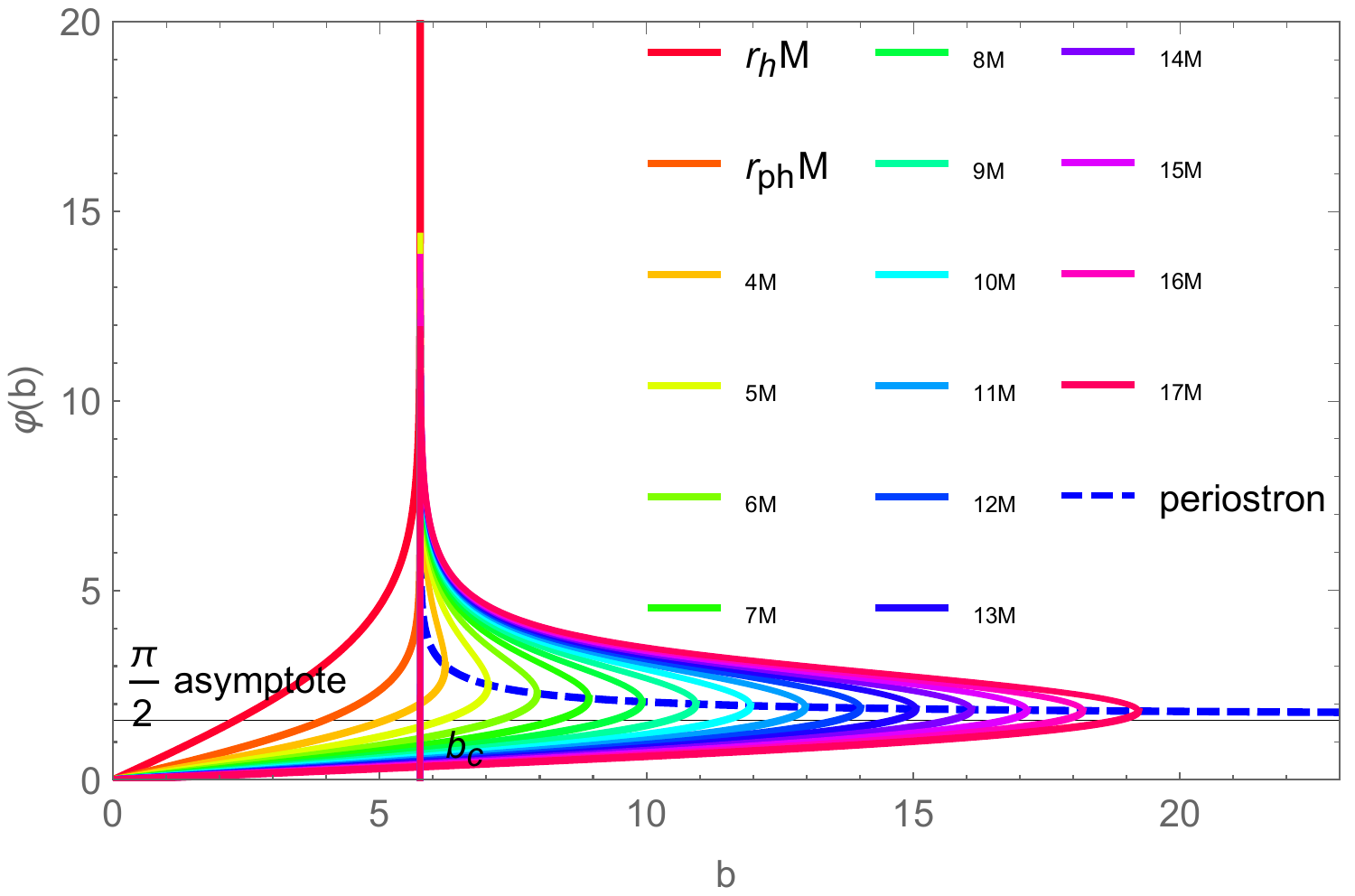}\hfill
    \includegraphics[width=0.5\textwidth]{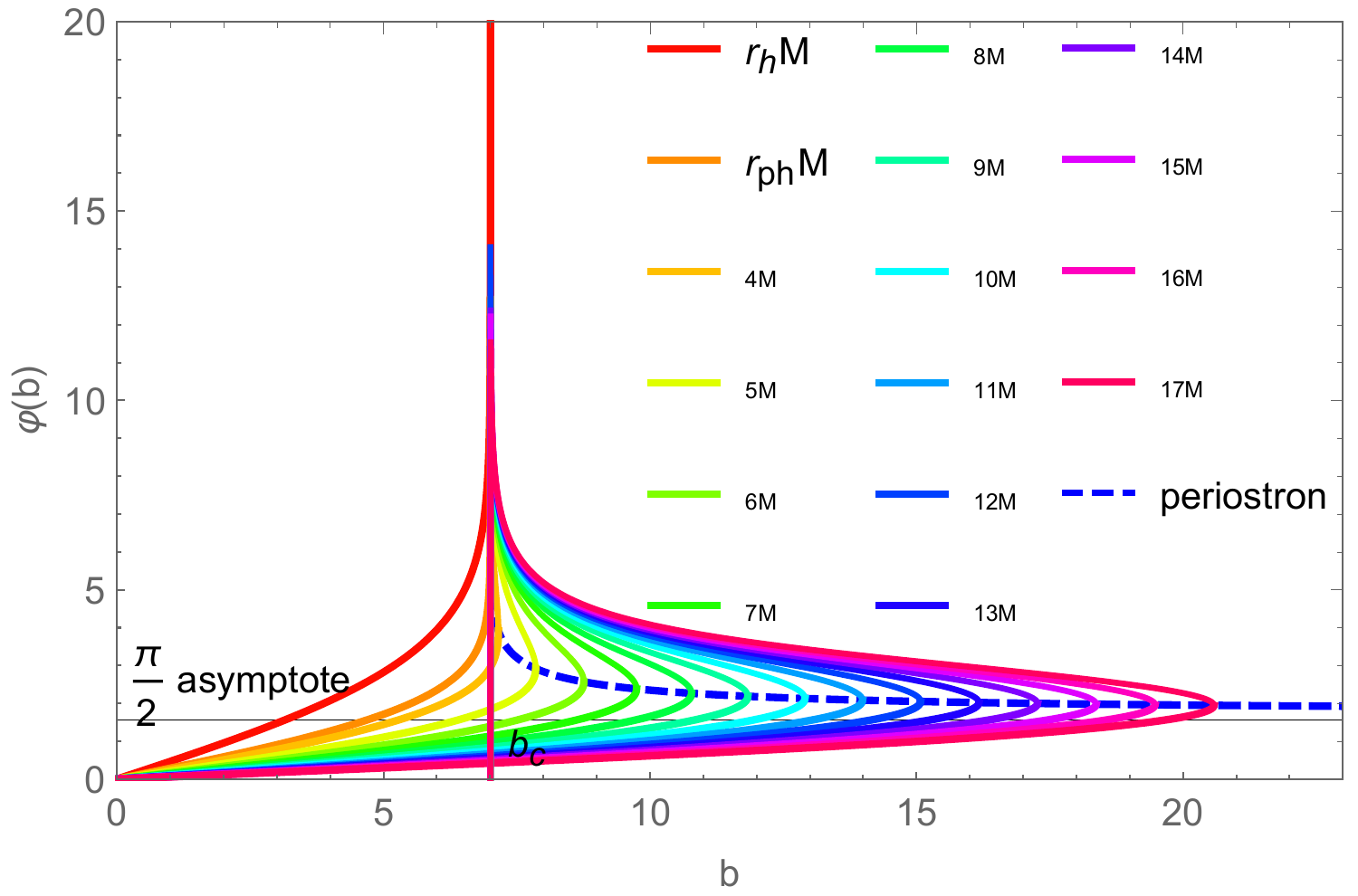}
\caption{Deflection angle $\varphi(b)$ corresponding to intersections as a function of $b$ for different $r$, with $l=0.9,a=0.1$ (left); $l=0.9,a=0.2$ (right).}
\label{fig:2} 
\end{figure*}

The impact parameters of black holes with different parameters are shown in Table \ref{tab:1}.
From Table \ref{tab:1}, it can be seen that the photon sphere radius \( r_{\rm h} \),  the impact parameter \( b_{\rm c} \) and the innermost stable circular orbit \( r_{\rm isco} \)  decrease slowly as the parameter \( l \) increases, and increase significantly as the parameter \( a \) increases. The value of the impact parameter \( b_{\rm c} \) here also corresponds to the value of the black hole shadow radius. The EHT observed that the 3$\delta$ confidence interval of the shadow radius of the M87* black hole is in the range of $(2.546M \leq Rs\leq 7.846M)$~\cite{Pantig:2024rmr}. In Table 1, when \( l = 0.5 \) and \( a = 0.8 \), the black hole shadow radius \( b_{\rm c} = 58.0861 \), which is far beyond the upper limit of this range. In the subsequent research, we will focus on the parameter values within this observed interval, mainly studying the cases where \( a = 0 \), \( 0.1 \), and \( 0.2 \).

\subsection{Thin Accretion Disk Imaging}
In the previous subsection, we used data to quantitatively describe the parameters related to the black hole shadow. Here, we adopt the method proposed by You \textit{et al.} for simulating the primary and secondary images of the accretion disk in the observer’s coordinate system (see Ref.~\cite{You:2024uql,You:2024jeu} for details). In this coordinate system, the black hole is described in spherical coordinates $(r, \theta, \phi)$, and the observer is located at $(\infty, \theta, 0)$. The observer’s visual coordinate system is defined as $(b, \alpha)$. When the rotation angle of the photon propagation plane in the observer’s visual coordinate system is $\alpha$, the angle between the line connecting the intersection point of this plane with the constant-$r$ orbit of the black hole and the origin $O$ of the black hole’s spherical coordinate system, and the rotation axis of the $\alpha/(\alpha + \pi)$ plane, is given by:
\begin{equation}
\varphi = \frac{\pi}{2} + \arctan\left( \tan\theta \sin\alpha \right)
\label{equ:13}
\end{equation}

When the photon impact parameter \( b \) is close to \( b_c \), the trajectory of the photon bends more significantly. For the same source point \( Q \), the photons emitted will appear as multiple image points \( q_n \) in the observer's view, where the subscript n denotes the order, corresponding to the magnitude of the rotation angle \( \varphi \) experienced by the photon from the source point to the image point. 

\begin{equation}
\varphi^n = 
\begin{cases} 
\frac{n}{2} 2\pi + (-1)^n \left[ \frac{\pi}{2} + \arctan\left( \tan\theta \sin\alpha \right) \right], & \text n \text{ is even}, \\
\frac{n+1}{2} 2\pi + (-1)^n \left[ \frac{\pi}{2} + \arctan\left( \tan\theta \sin\alpha \right) \right], & \text n \text{ is odd},
\end{cases}
\label{equ:14}
\end{equation}

In spacetime, through the photon trajectories on the equatorial plane, one can calculate the total deflection angle \(\varphi\) when photons with different impact parameters intersect the equatorial circular orbits of the accretion disk at radius $r$

\begin{equation}
\varphi(b) = \int_{0}^{u_r} \frac{1}{\sqrt{G(u)}} du
\label{equ:15}
\end{equation}

 \begin{figure*}[t]
\centering
\begin{minipage}{0.3\textwidth}
    \includegraphics[width=\linewidth]{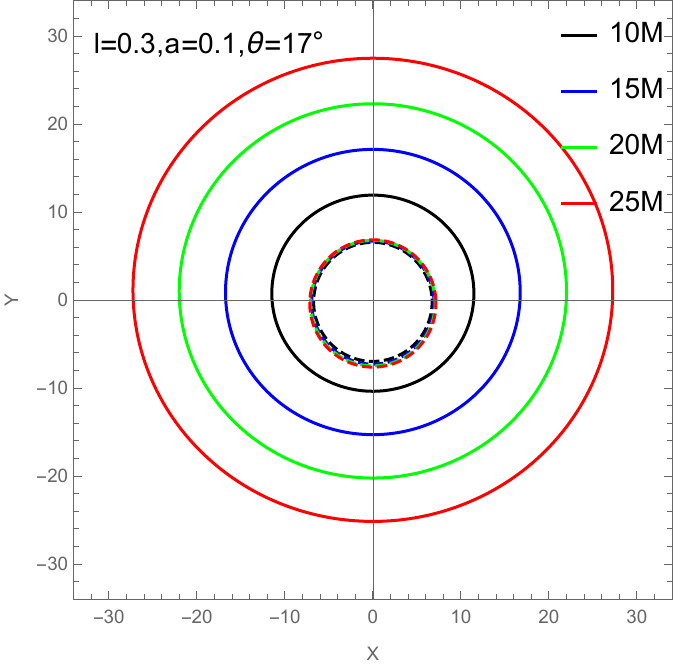}
\end{minipage}
\begin{minipage}{0.3\textwidth}
    \includegraphics[width=\linewidth]{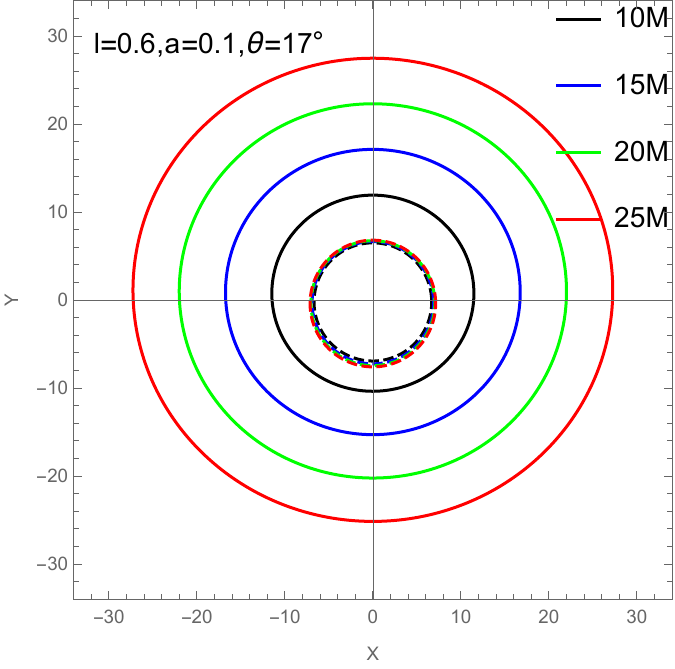}
\end{minipage}
\begin{minipage}{0.3\textwidth}
    \includegraphics[width=\linewidth]{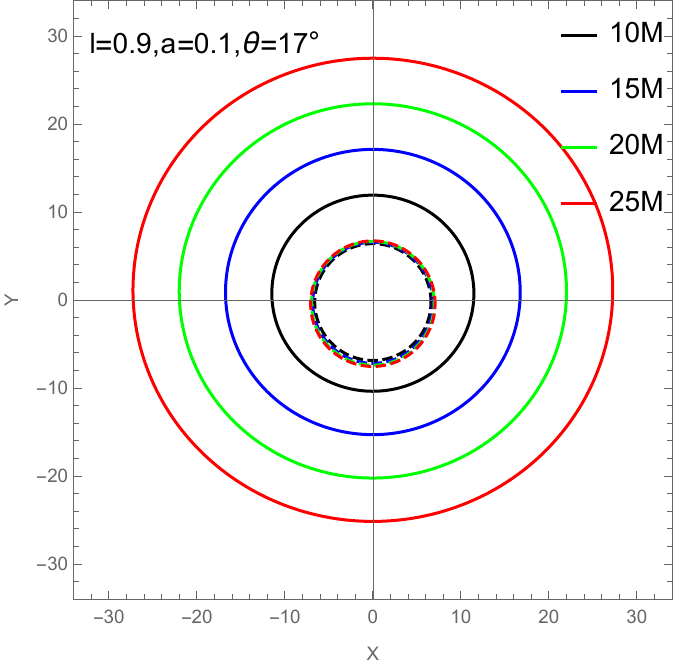}
\end{minipage}

\begin{minipage}{0.3\textwidth}
    \includegraphics[width=\linewidth]{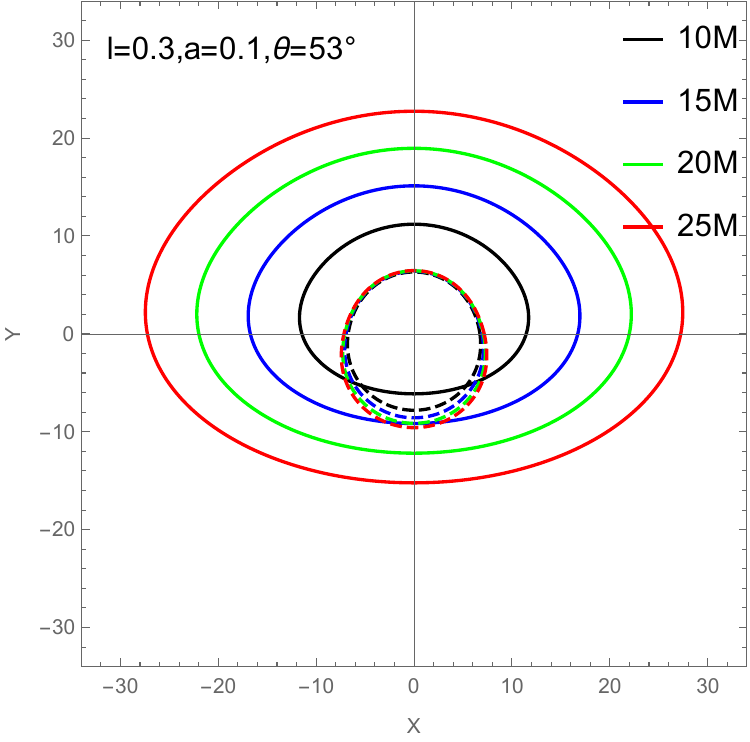}
\end{minipage}
\begin{minipage}{0.3\textwidth}
    \includegraphics[width=\linewidth]{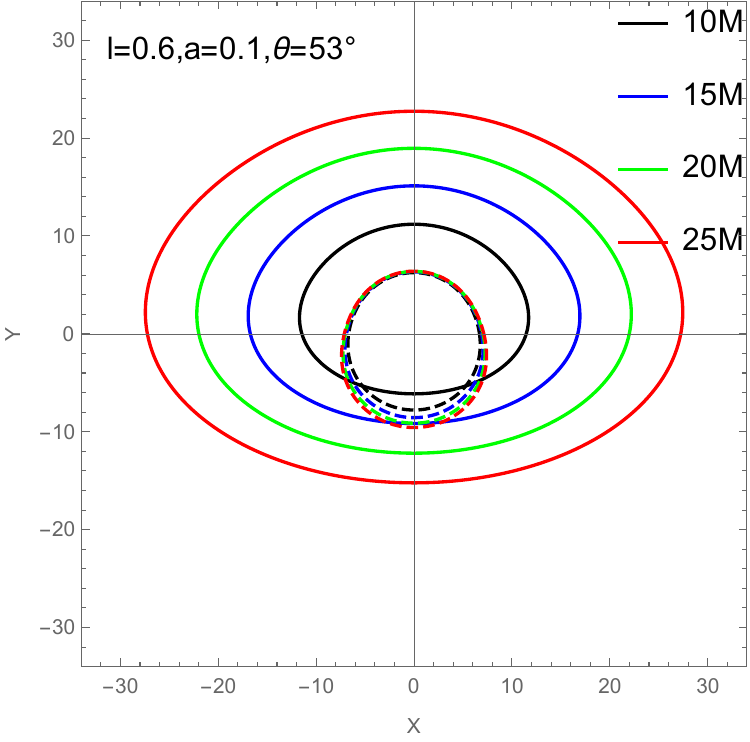}
\end{minipage}
\begin{minipage}{0.3\textwidth}
    \includegraphics[width=\linewidth]{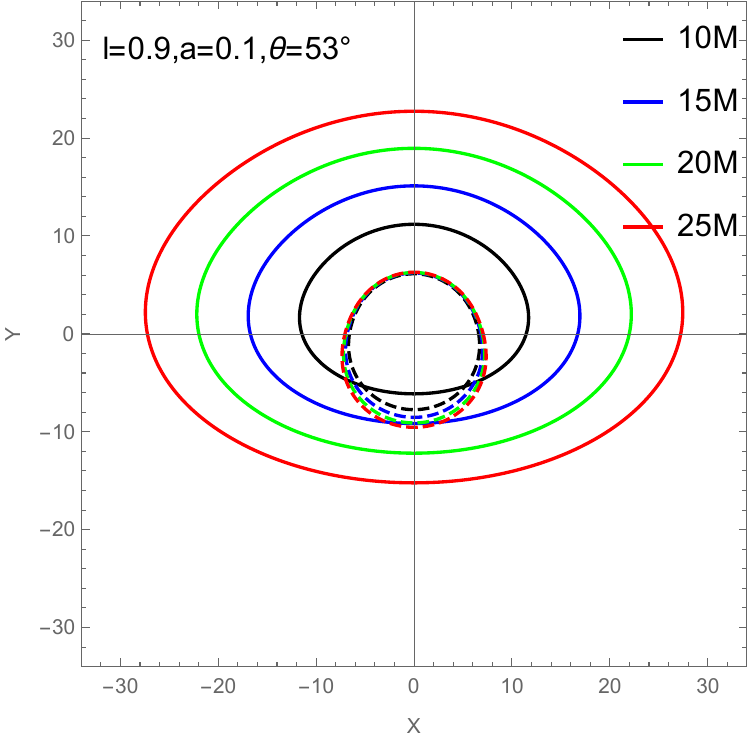}
\end{minipage}

\begin{minipage}{0.3\textwidth}
    \includegraphics[width=\linewidth]{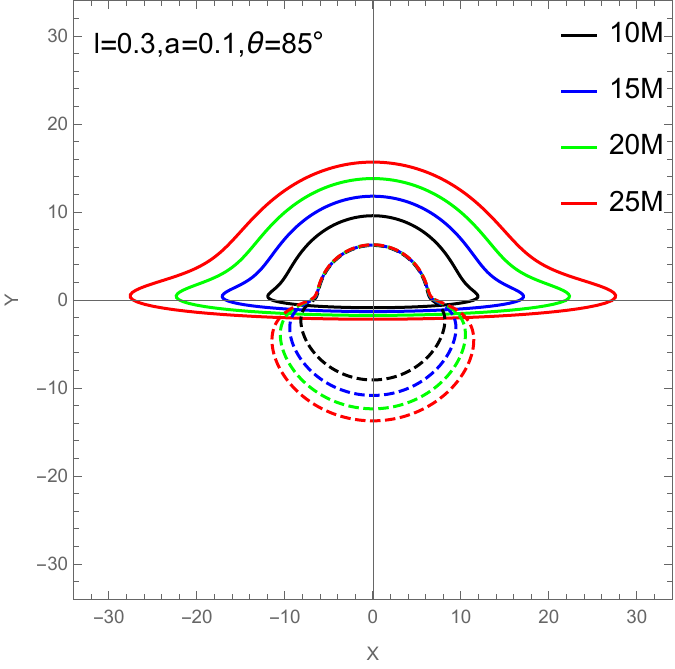}
\end{minipage}
\begin{minipage}{0.3\textwidth}
    \includegraphics[width=\linewidth]{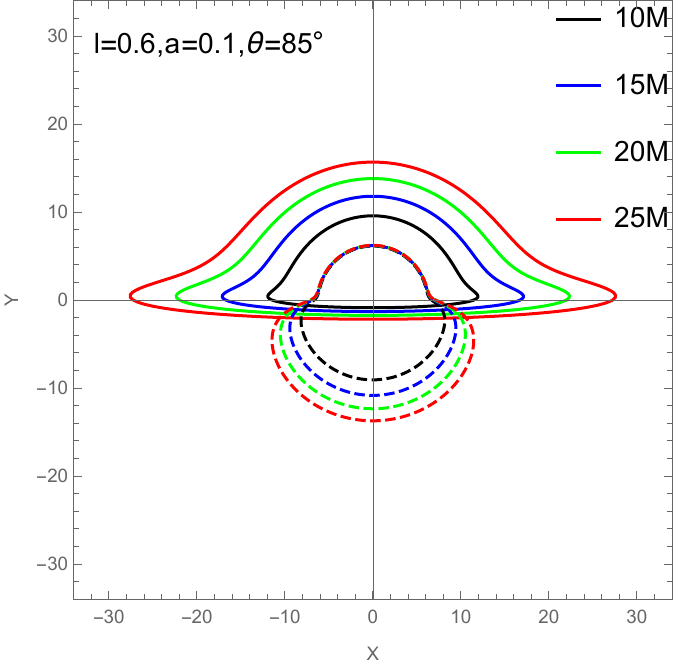}
\end{minipage}
\begin{minipage}{0.3\textwidth}
    \includegraphics[width=\linewidth]{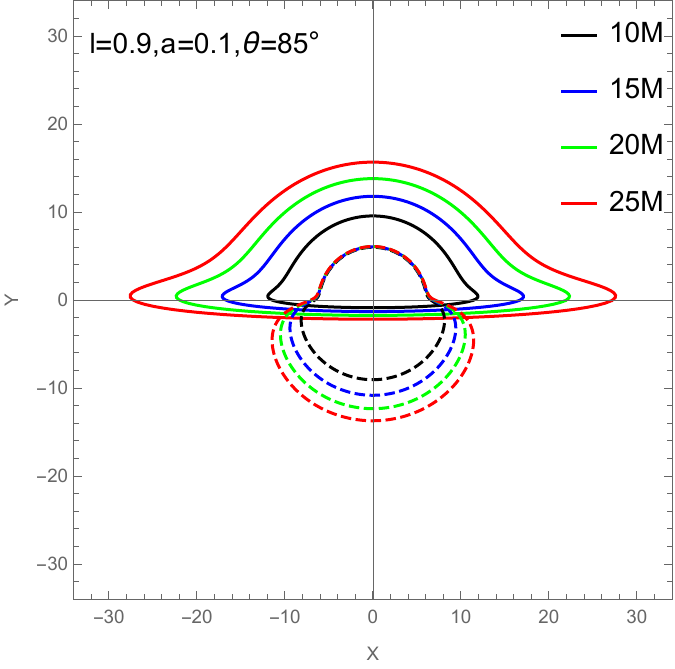}
\end{minipage}

\caption{Direct and indirect images of equal-r orbits of accretion disks under different parameters.}
\centering
\label{fig:3}
\end{figure*}

Fig. \ref{fig:2} shows the function graph of \( \varphi(b) \), where curves of different colors represent different equal-\( r \) orbits. Each point \( (b, \varphi) \) on the curves indicates the deflection angle \( \varphi \) corresponding to a photon with impact parameter \( b \) when it reaches the equal-\( r \) orbit. The blue dashed line represents the deflection angles of photons with different impact parameters when they reach the perihelion. Let the blue dashed line be denoted as \( \varphi_{(\rm blue)}(b) \), the curves below the blue dashed line as \( \varphi_{(\rm down)}(b) \), and those above as \( \varphi_{(\rm up)}(b) \). These can be expressed as:

\begin{equation}
\varphi_{(\rm blue)}(b)= \int_{0}^{u_{\text{min}}} \frac{1}{\sqrt{G(u)}} du
\label{equ:16}
\end{equation}

\begin{equation}
\varphi_{(\rm down)}(b)=  \int_{0}^{u_r} \frac{1}{\sqrt{G(u)}} du
\label{equ:17}
\end{equation}

\begin{equation}
\varphi_{\rm up}(b) = 2 \int_{0}^{u_{\text{min}}} \frac{1}{\sqrt{G(u)}} du - \int_{0}^{u_r} \frac{1}{\sqrt{G(u)}} du
\label{equ:18}
\end{equation}

\begin{figure*}[t]
\centering
\begin{minipage}{0.3\textwidth}
    \includegraphics[width=\linewidth]{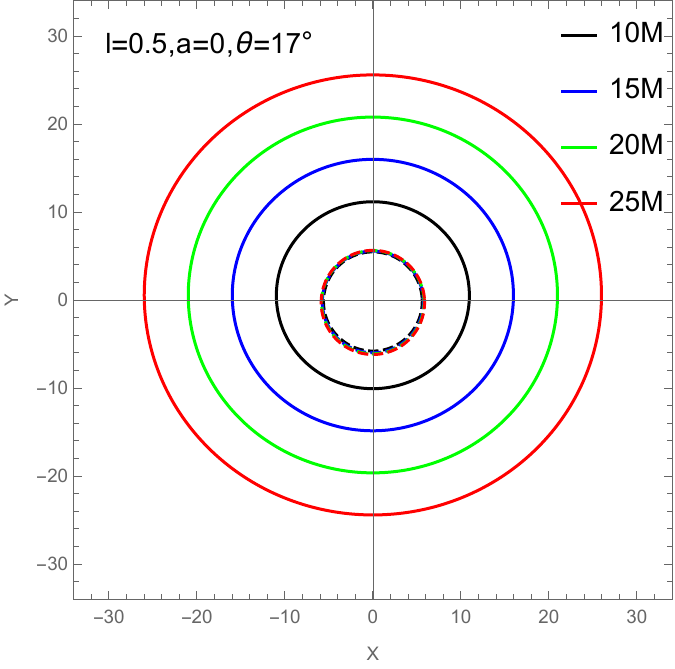}
\end{minipage}
\begin{minipage}{0.3\textwidth}
    \includegraphics[width=\linewidth]{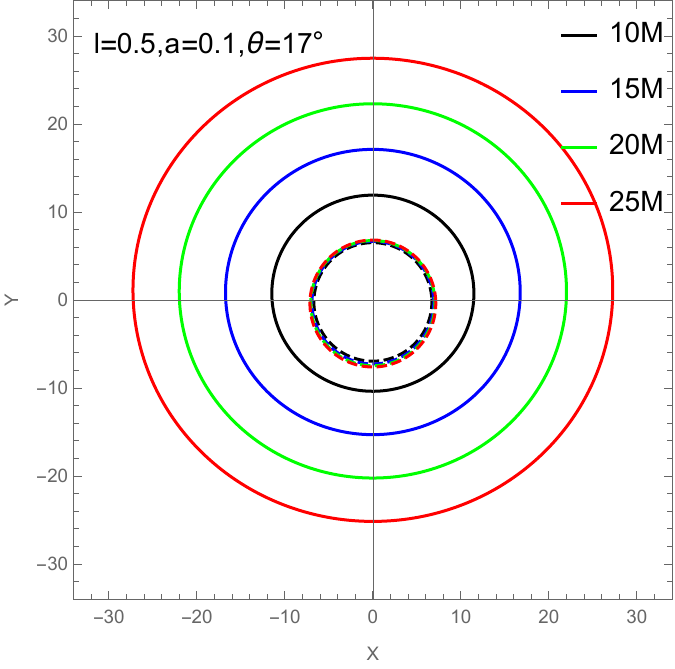}
\end{minipage}
\begin{minipage}{0.3\textwidth}
    \includegraphics[width=\linewidth]{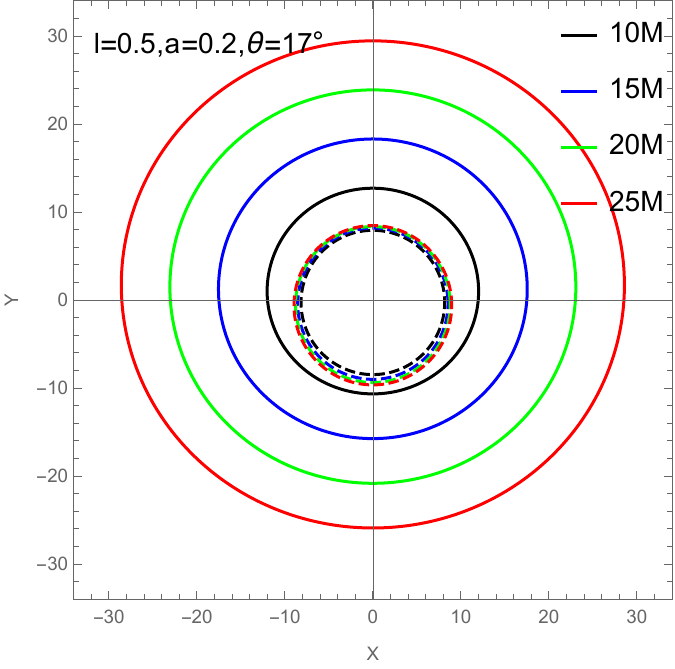},
\end{minipage}

\begin{minipage}{0.3\textwidth}
    \includegraphics[width=\linewidth]{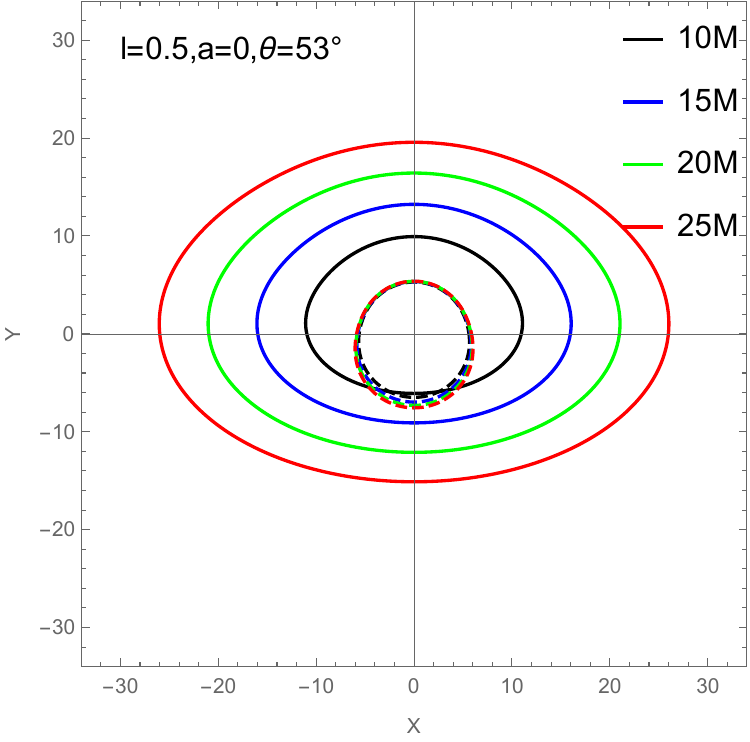}
\end{minipage}
\begin{minipage}{0.3\textwidth}
    \includegraphics[width=\linewidth]{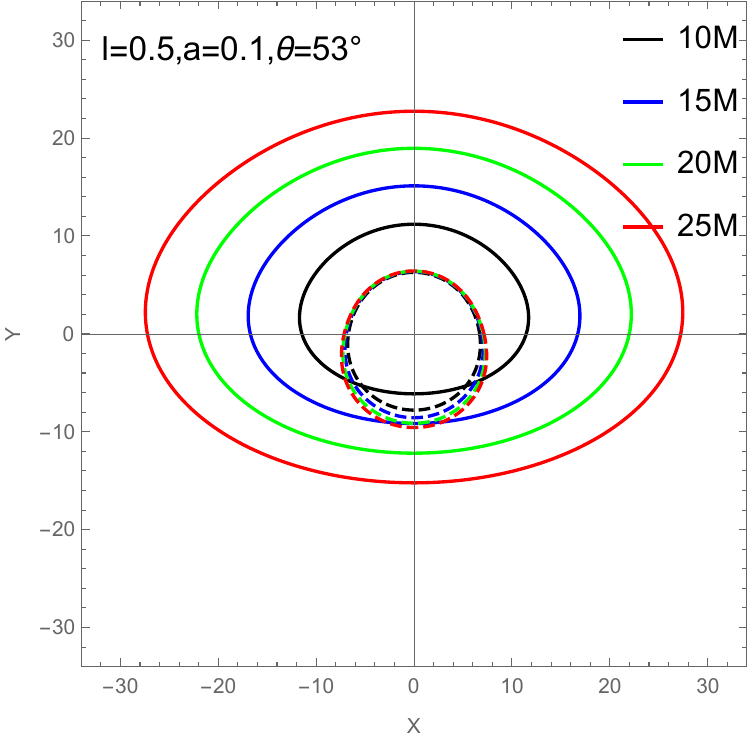}
\end{minipage}
\begin{minipage}{0.3\textwidth}
    \includegraphics[width=\linewidth]{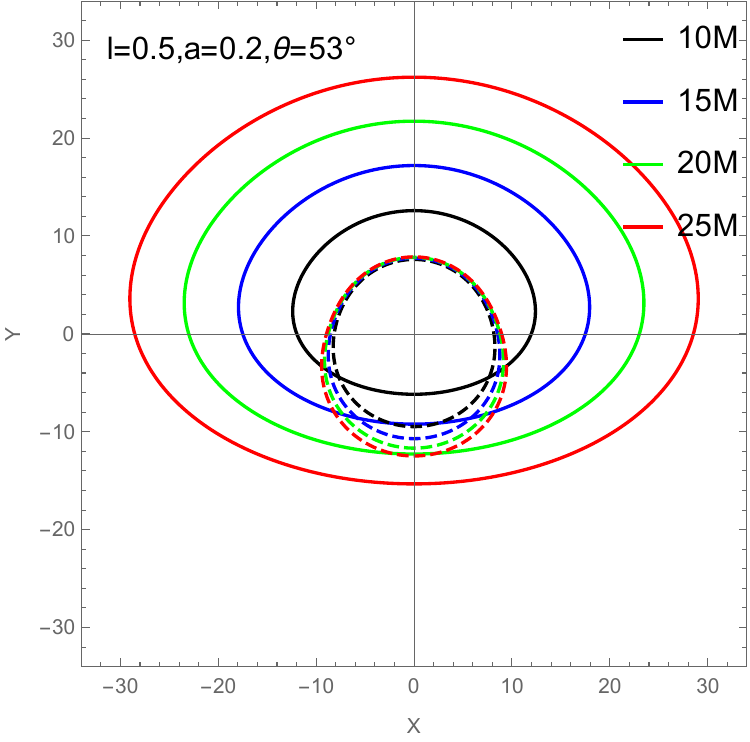}
\end{minipage}

\begin{minipage}{0.3\textwidth}
    \includegraphics[width=\linewidth]{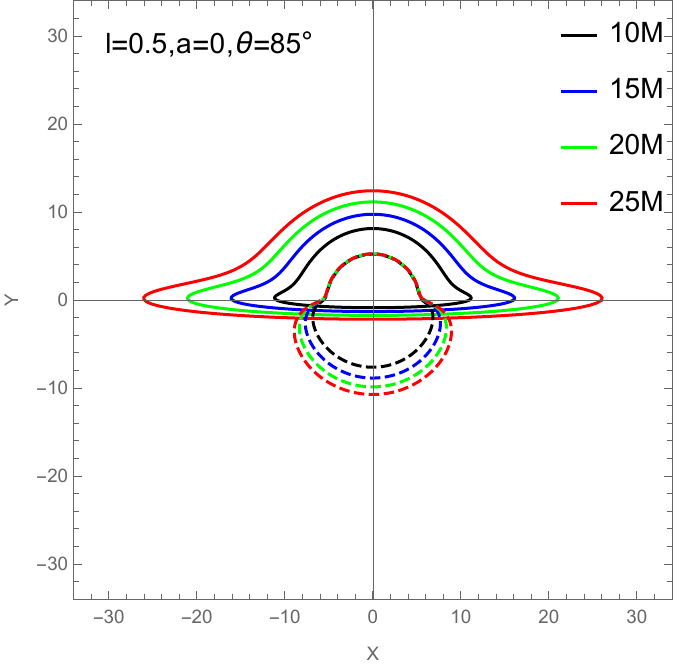}
\end{minipage}
\begin{minipage}{0.3\textwidth}
    \includegraphics[width=\linewidth]{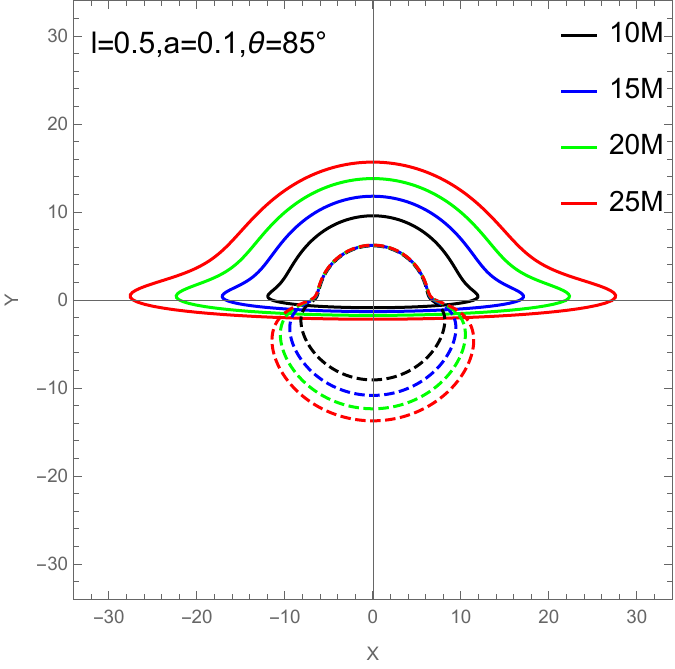}
\end{minipage}
\begin{minipage}{0.3\textwidth}
    \includegraphics[width=\linewidth]{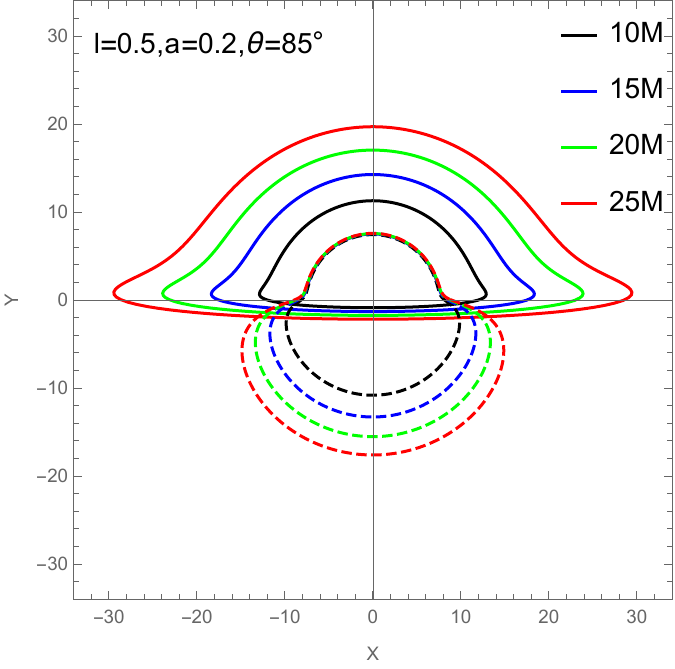}
\end{minipage}

\caption{Direct and indirect images of equal-r orbits of accretion disks under different parameters.}
\label{fig:4}
\end{figure*}

\begin{figure*}[t]
\centering
\begin{minipage}{0.3\textwidth}
    \includegraphics[width=\linewidth]{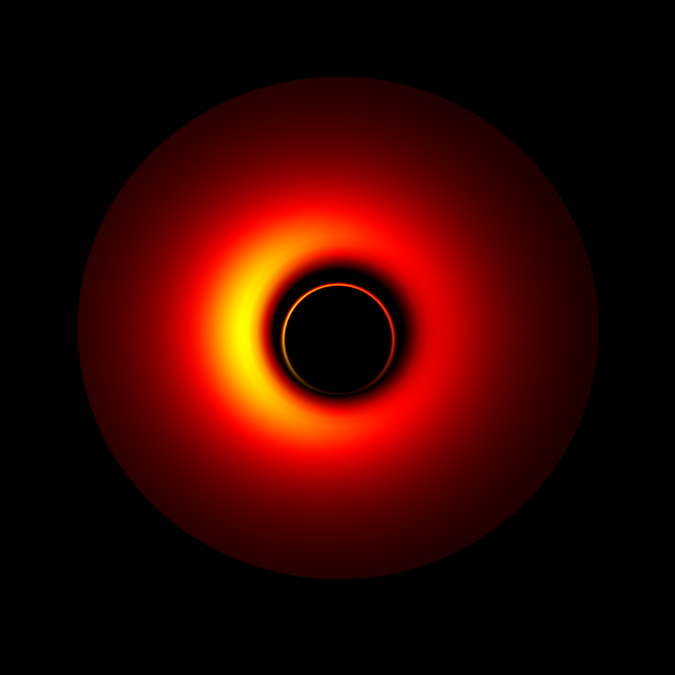}
    \centering $l=0.3, a=0.1, \theta=17^\circ$
\end{minipage}
\begin{minipage}{0.3\textwidth}
    \includegraphics[width=\linewidth]{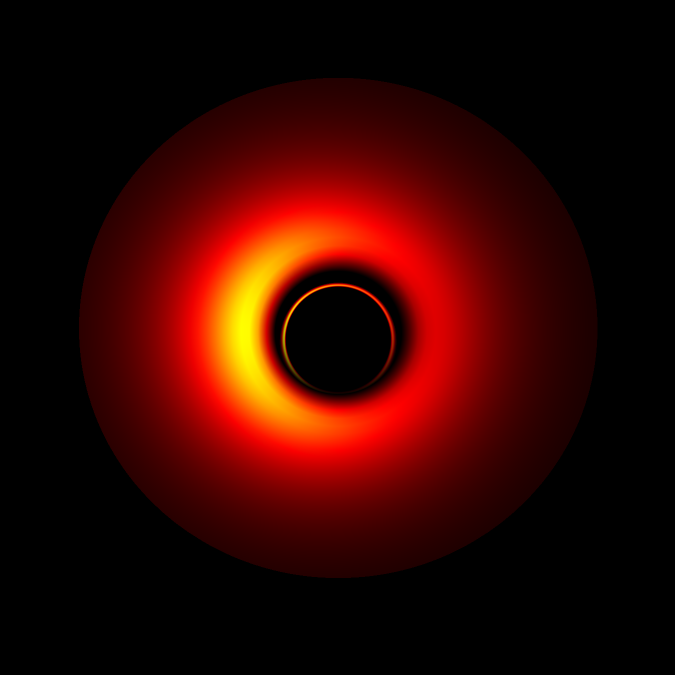}
    \centering $l=0.6, a=0.1, \theta=17^\circ$
\end{minipage}
\begin{minipage}{0.3\textwidth}
    \includegraphics[width=\linewidth]{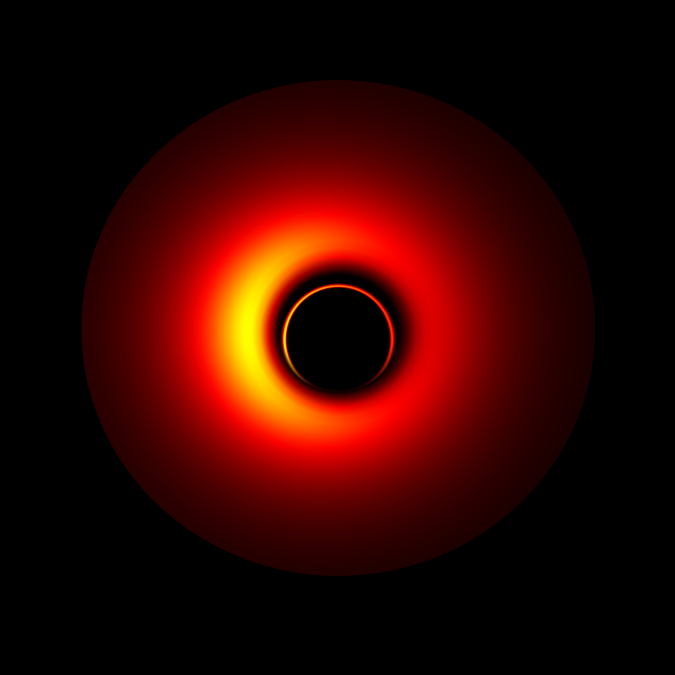}
    \centering $l=0.9, a=0.1, \theta=17^\circ$
\end{minipage}

\begin{minipage}{0.3\textwidth}
    \includegraphics[width=\linewidth]{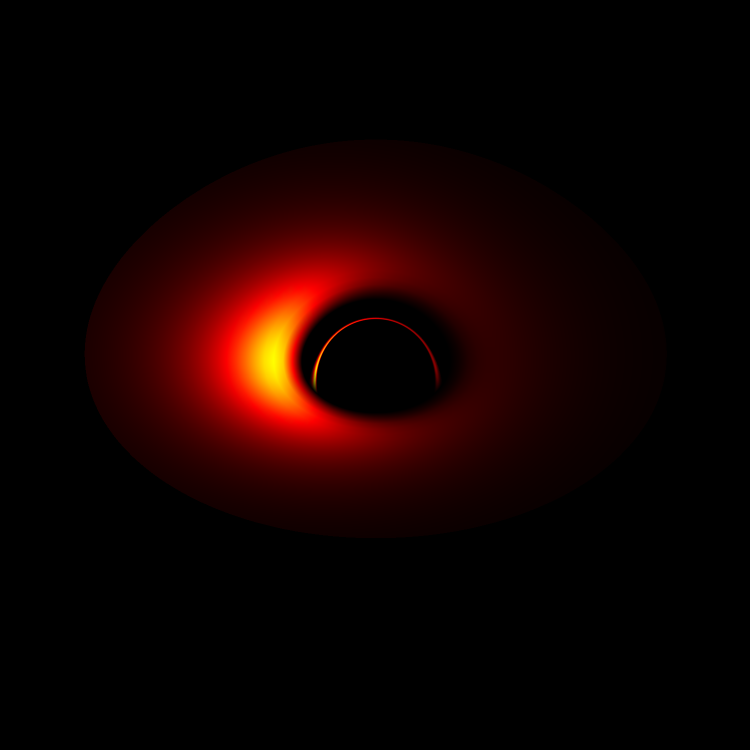}
    \centering $l=0.3, a=0.1, \theta=53^\circ$
\end{minipage}
\begin{minipage}{0.3\textwidth}
    \includegraphics[width=\linewidth]{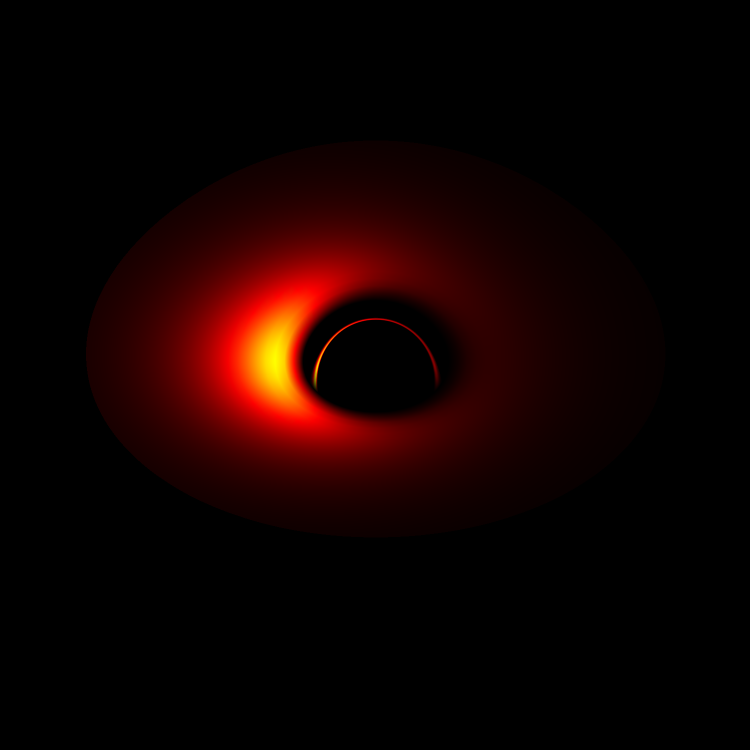}
    \centering $l=0.6, a=0.1, \theta=53^\circ$
\end{minipage}
\begin{minipage}{0.3\textwidth}
    \includegraphics[width=\linewidth]{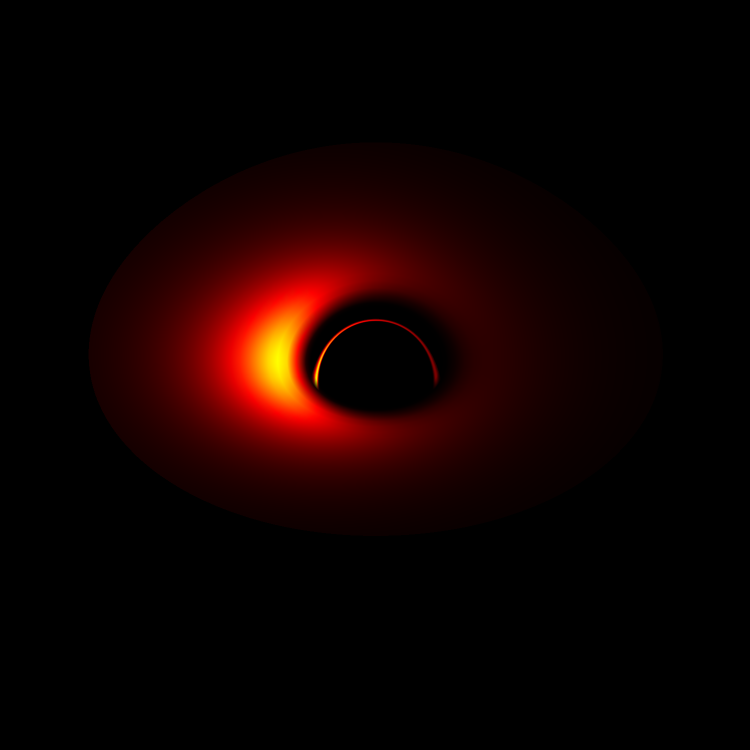}
    \centering $l=0.9, a=0.1, \theta=53^\circ$
\end{minipage}

\begin{minipage}{0.3\textwidth}
    \includegraphics[width=\linewidth]{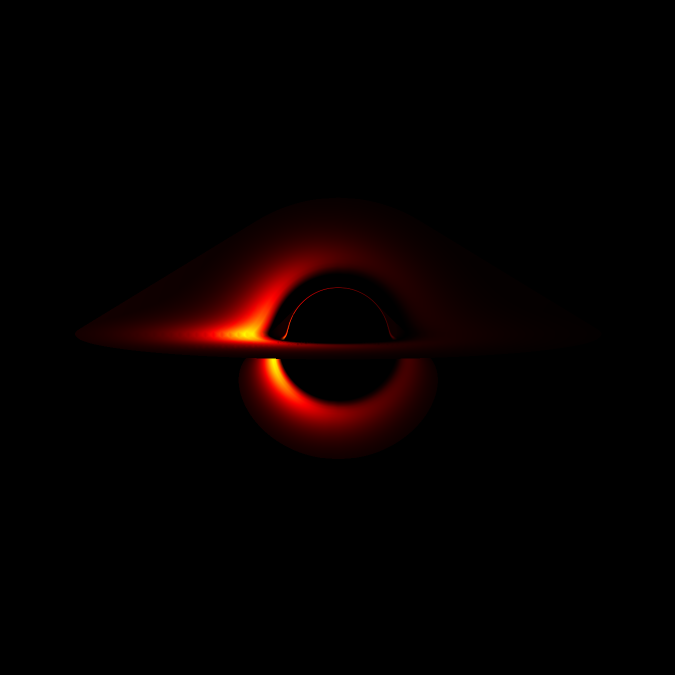}
    \centering $l=0.3, a=0.1, \theta=85^\circ$
\end{minipage}
\begin{minipage}{0.3\textwidth}
    \includegraphics[width=\linewidth]{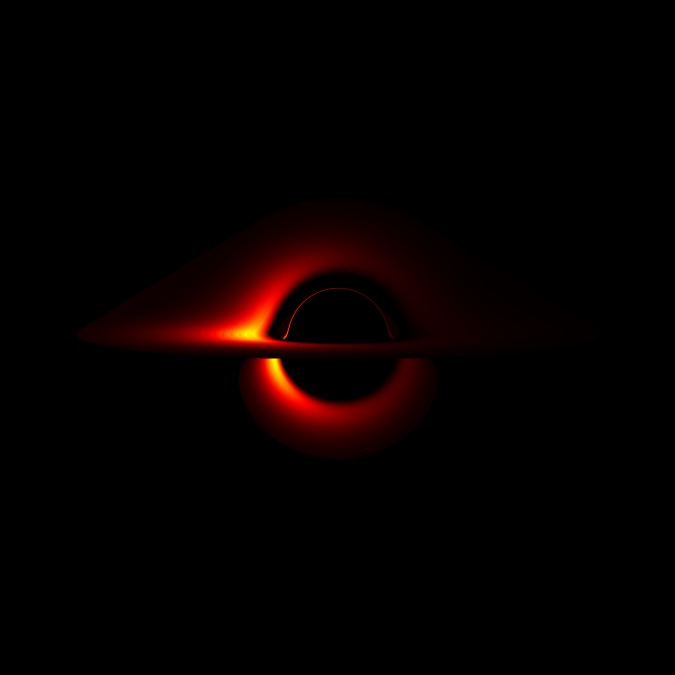}
    \centering $l=0.6, a=0.1, \theta=85^\circ$
\end{minipage}
\begin{minipage}{0.3\textwidth}
    \includegraphics[width=\linewidth]{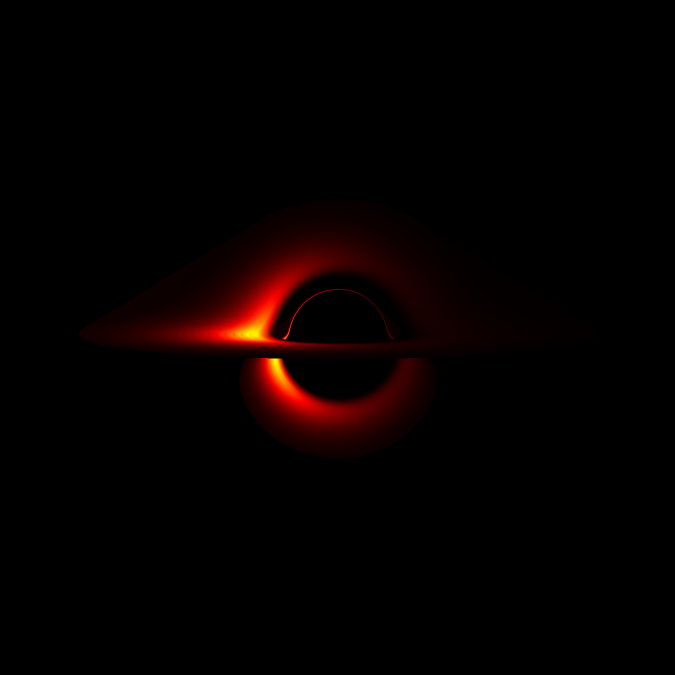}
    \centering $l=0.9, a=0.1, \theta=85^\circ$
\end{minipage}

\caption{Direct and indirect intensity images of accretion disks under different parameters.}
\label{fig:5}
\end{figure*}

\begin{figure*}[t]
\centering
\begin{minipage}{0.3\textwidth}
    \includegraphics[width=\linewidth]{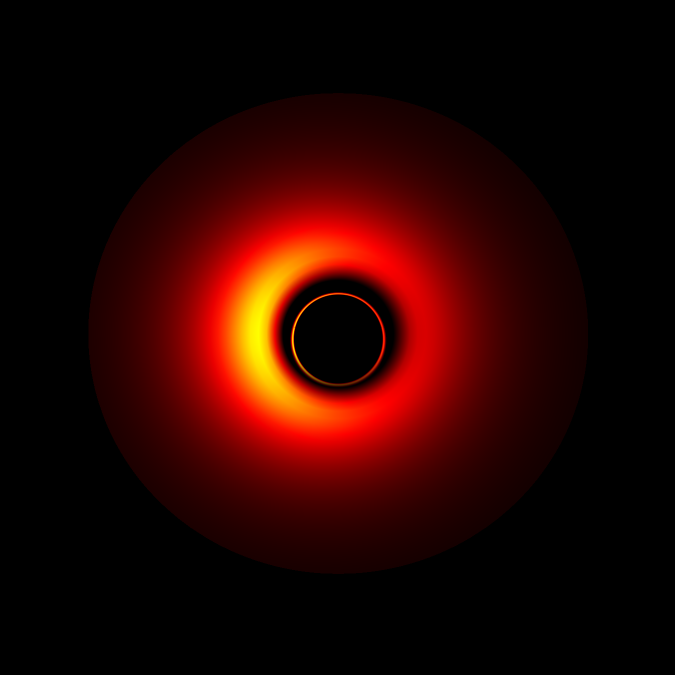}
    \centering $l=0.5, a=0, \theta=17^\circ$
\end{minipage}
\begin{minipage}{0.3\textwidth}
    \includegraphics[width=\linewidth]{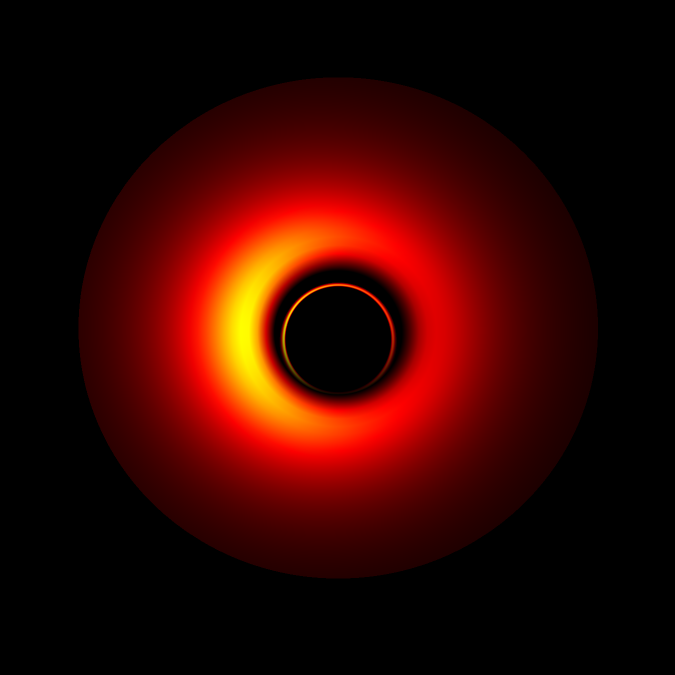}
    \centering $l=0.5, a=0.1, \theta=17^\circ$
\end{minipage}
\begin{minipage}{0.3\textwidth}
    \includegraphics[width=\linewidth]{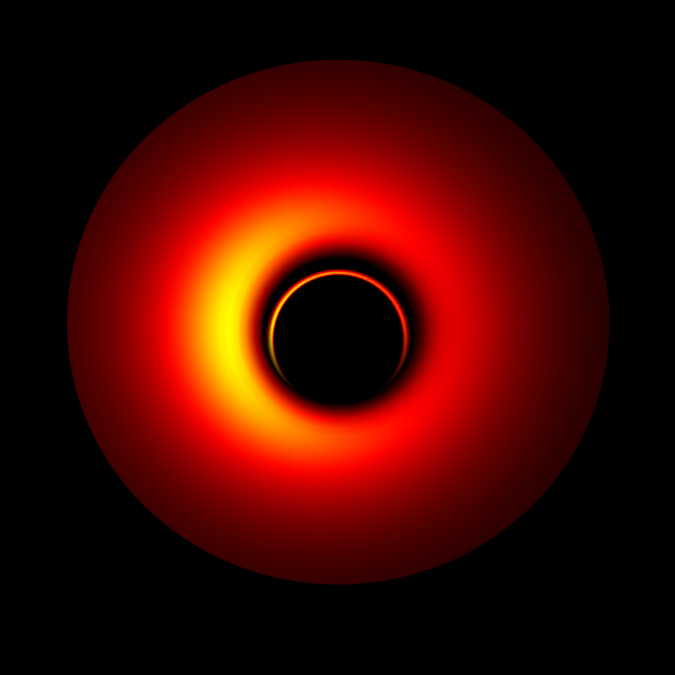}
    \centering $l=0.5, a=0.2, \theta=17^\circ$
\end{minipage}

\begin{minipage}{0.3\textwidth}
    \includegraphics[width=\linewidth]{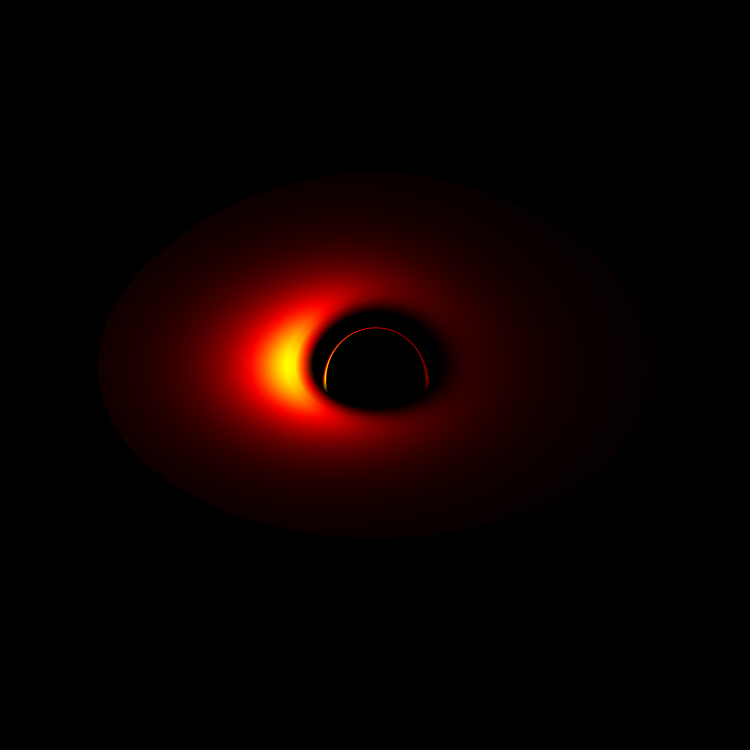}
    \centering $l=0.5, a=0, \theta=53^\circ$
\end{minipage}
\begin{minipage}{0.3\textwidth}
    \includegraphics[width=\linewidth]{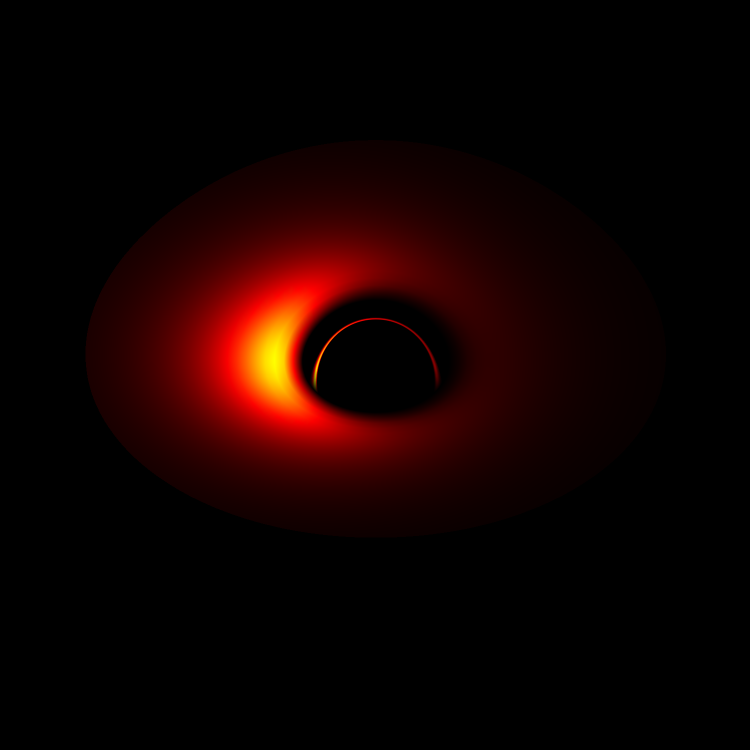}
    \centering $l=0.5, a=0.1, \theta=53^\circ$
\end{minipage}
\begin{minipage}{0.3\textwidth}
    \includegraphics[width=\linewidth]{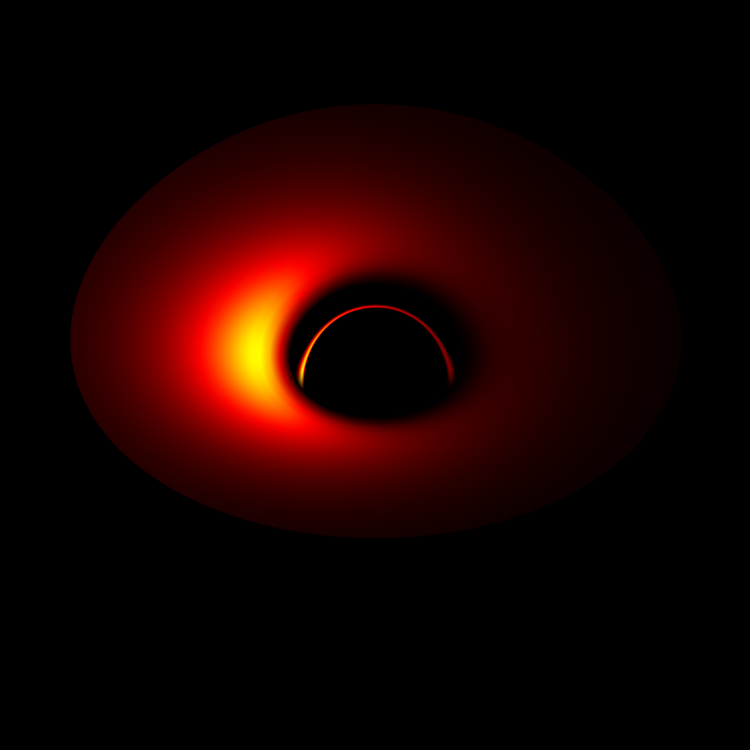}
    \centering $l=0.5, a=0.2, \theta=53^\circ$
\end{minipage}

\begin{minipage}{0.3\textwidth}
    \includegraphics[width=\linewidth]{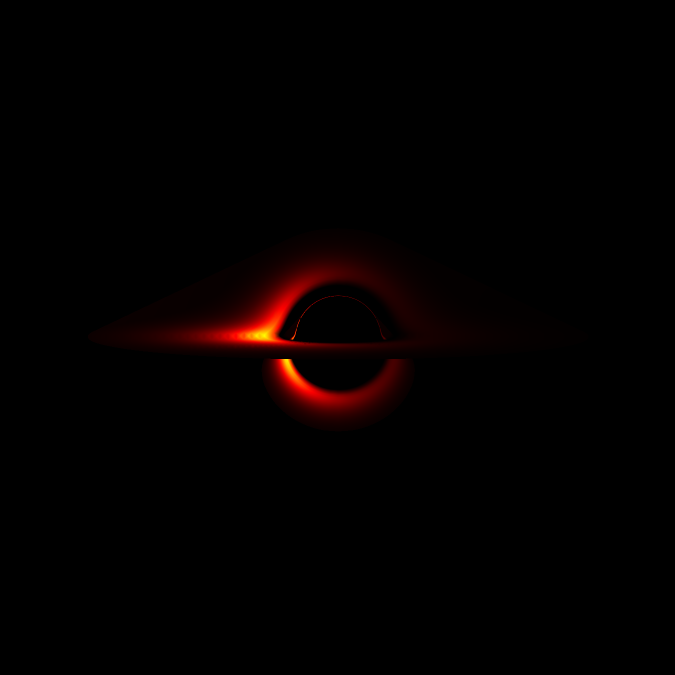}
    \centering $l=0.5, a=0, \theta=85^\circ$
\end{minipage}
\begin{minipage}{0.3\textwidth}
    \includegraphics[width=\linewidth]{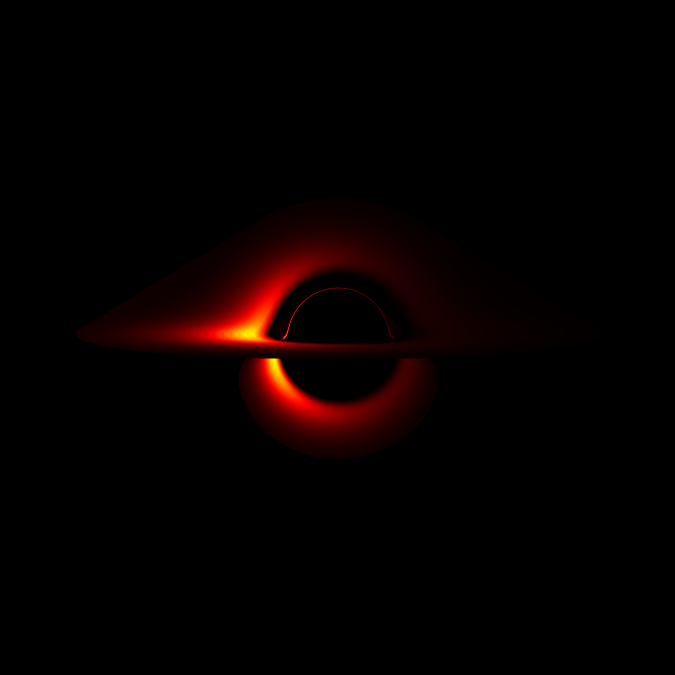}
    \centering $l=0.5, a=0.1, \theta=85^\circ$
\end{minipage}
\begin{minipage}{0.3\textwidth}
    \includegraphics[width=\linewidth]{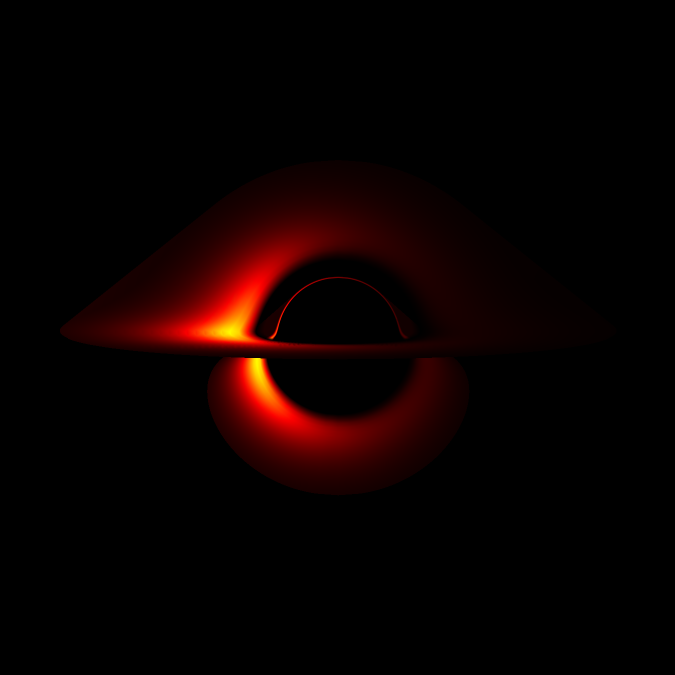}
    \centering $l=0.5, a=0.2, \theta=85^\circ$
\end{minipage}

\caption{Direct and indirect intensity images of accretion disks under different parameters.}
\label{fig:6}
\end{figure*}

\begin{figure*}[t]
\centering
  \includegraphics[width=0.5\textwidth]{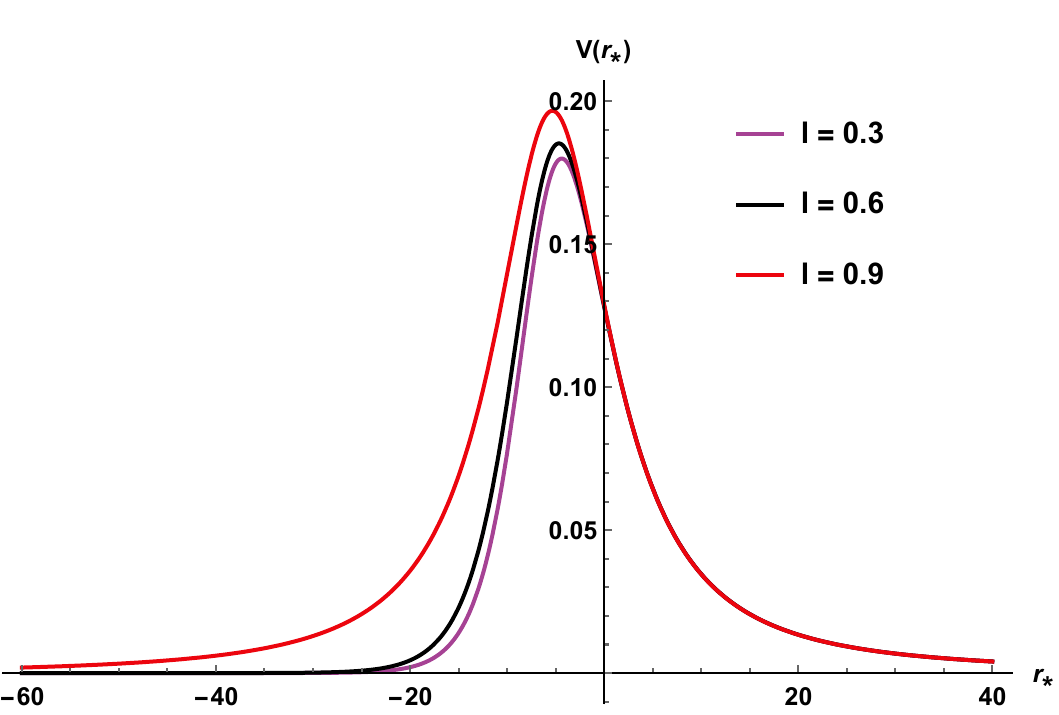}\hfill
    \includegraphics[width=0.5\textwidth]{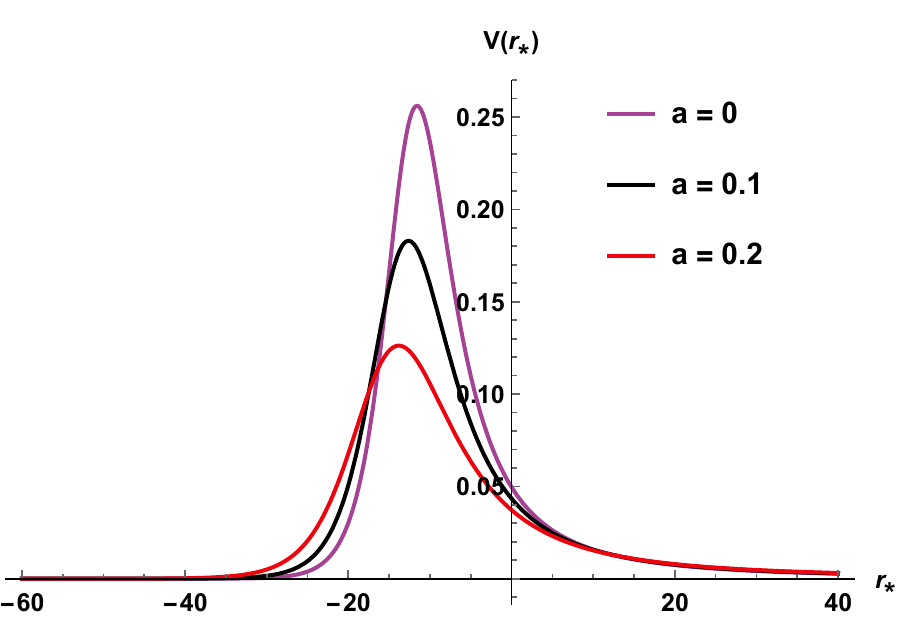}
\caption{The effective potential of the black hole perturbed by  scalar field under different parameter values, with $a = 0.1$ (left) and $l = 0.5$ (right).}
\label{fig:7}       % Give a unique label
\end{figure*}

\begin{figure*}[t]
\centering
  \includegraphics[width=0.5\textwidth]{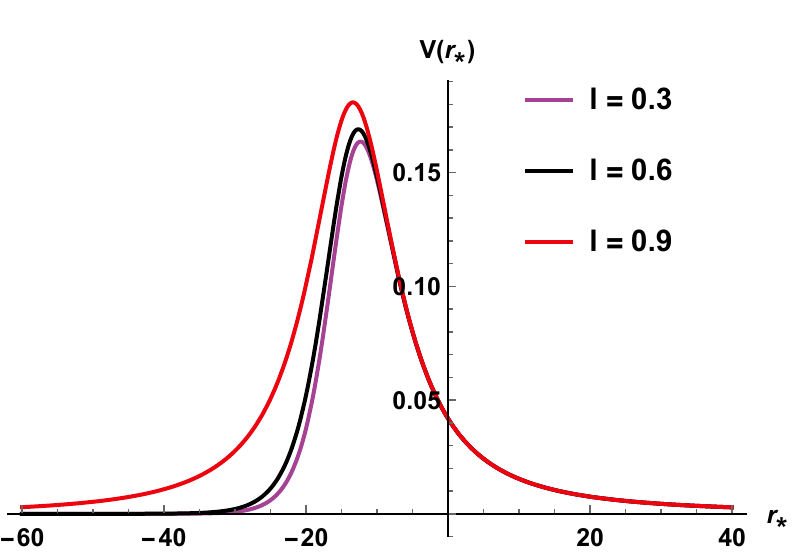}\hfill
    \includegraphics[width=0.5\textwidth]{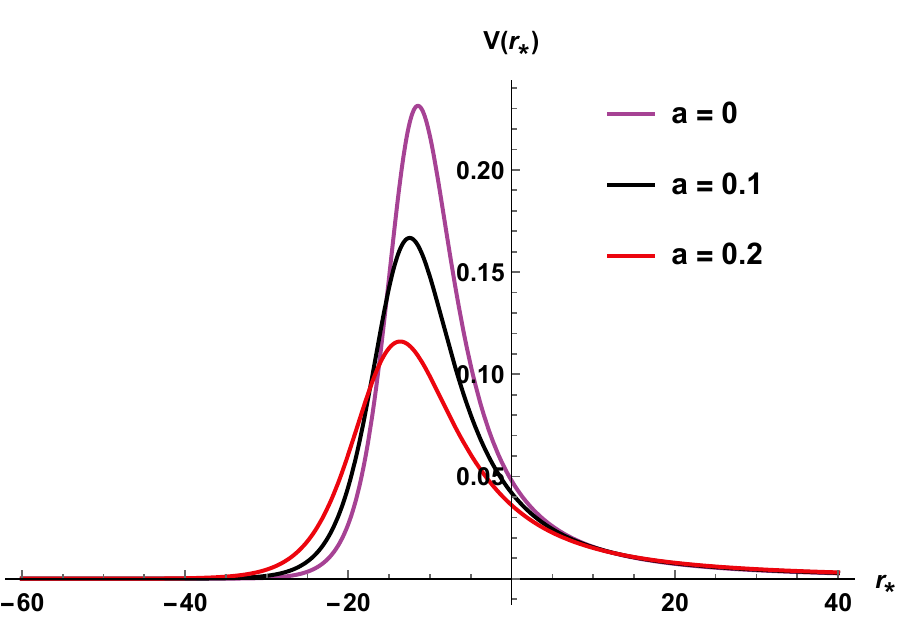}
\caption{The effective potential of the black hole perturbed by electromagnetic field under different parameter values, with $a = 0.1$ (left) and $l = 0.5$ (right)}
\label{fig:8}       % Give a unique label
\end{figure*}

\begin{figure*}[t]
\centering
  \includegraphics[width=0.5\textwidth]{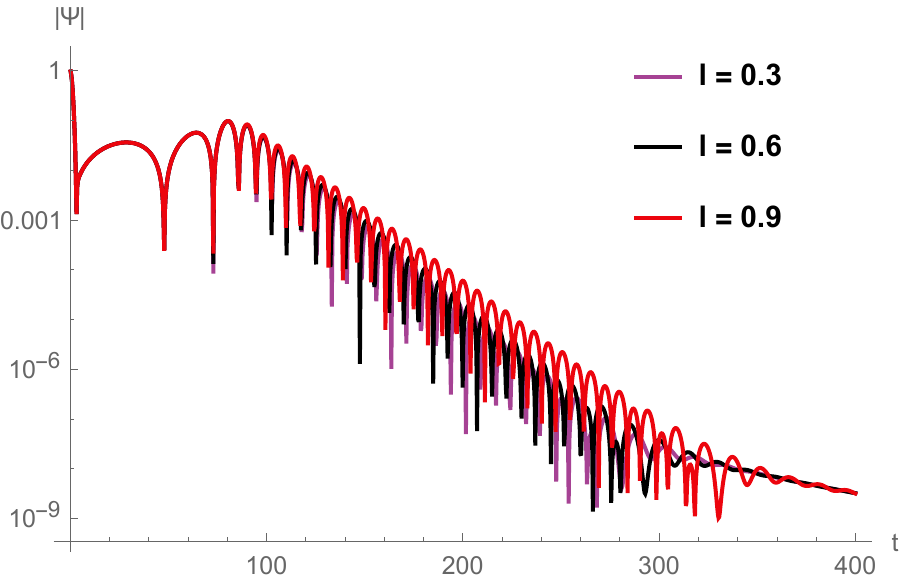}\hfill
    \includegraphics[width=0.5\textwidth]{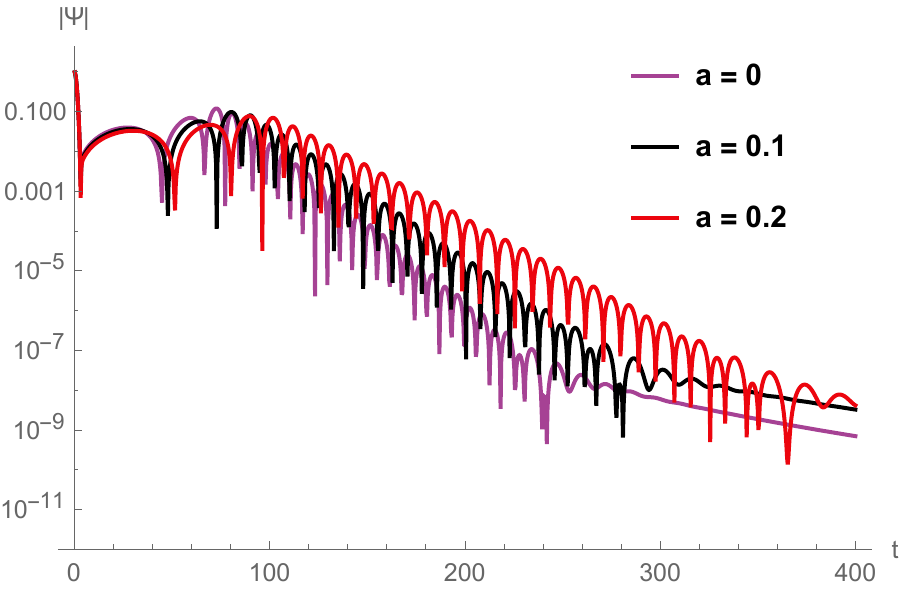}

\caption{The time-domain profile diagrams of the black holes perturbed by scalar fields under different parameters, $a = 0.1$ (left) and $l = 0.5$ (right).}
\label{fig:9}    
\end{figure*}

\begin{figure*}[t]
\centering
  \includegraphics[width=0.5\textwidth]{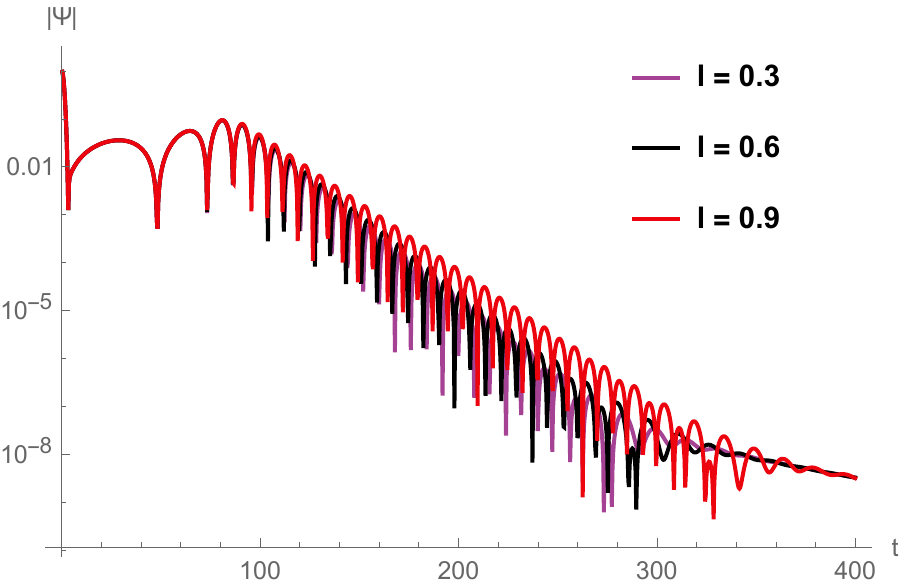}\hfill
    \includegraphics[width=0.5\textwidth]{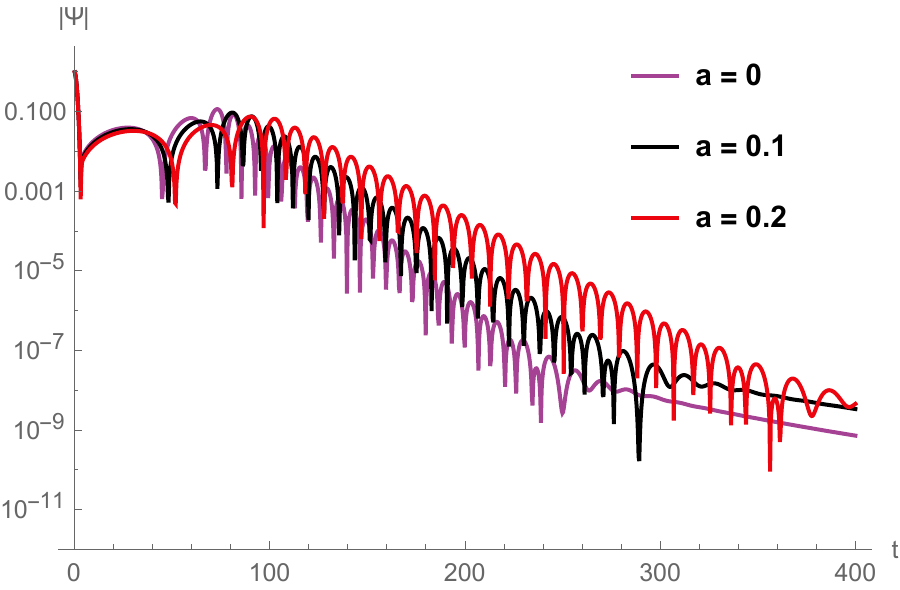}
\caption{The time-domain profile diagrams of the black holes perturbed by electromagnetic field under different parameters, $a = 0.1$ (left) and $l = 0.5$ (right)}
\label{fig:10}       % Give a unique label
\end{figure*}

By solving Eqs. \ref{equ:13}, \ref{equ:17}, and \ref{equ:18}, we can plot the equal-$r$ curves in the observer’s visual plane, as shown in Figs. \ref{fig:3} and \ref{fig:4}. The first to third rows correspond to observation angles of $17^\circ$, $53^\circ$, and $85^\circ$, respectively, while the first to third columns correspond to increasing values of the parameter $l/a$. The solid lines represent the primary images, and the dashed lines denote the secondary images. To visualize the black hole imaging more intuitively, the intensity map of the accretion disk can be calculated using the Novikov--Thorne model. The radiation intensity emitted from the accretion disk at radius $r$ is given by~\cite{Page:1974he}:

\begin{equation}
F(r) = -\frac{\mathcal{M}\Omega'}{4\pi\sqrt{-g}(E - \Omega L)^2} \int_{r_{isco}}^{r} (E - \Omega L) L' dr
\label{equ:19}
\end{equation}
where $\mathcal{M}$ is the mass accretion rate of the disk, $\Omega$ is the angular velocity of the particles, and the prime symbol ``$'$'' denotes differentiation with respect to $r$. Considering the gravitational redshift effect, the radiation intensity formula is rewritten as:

\begin{equation}
1 + z = \frac{E_r}{E_\infty}
\label{equ:20}
\end{equation}

\begin{equation}
F_{\text{obs}} = \frac{F(r)}{(1+z)^4}
\label{equ:21}
\end{equation}

where the $z$ is the redshift factor, $E_r$ is the photon energy measured at the emission point of the particle, and $E_\infty$ is the photon energy measured by a static observer at infinity. Substituting Eqs. \ref{equ:19} and \ref{equ:20} into Eq. \ref{equ:21} yields:
\begin{equation}
F_{\rm obs} = \frac{-\frac{\mathcal{M}\Omega'}{4\pi\sqrt{-g}(E - \Omega L)^2} \int_{r_{in}}^{r} (E - \Omega L) L' dr}{\left( \frac{(1 + b \sin\theta \cos\alpha \Omega)}{\sqrt{-g_{tt} - g_{\phi\phi}\Omega^2}} \right)^4}
\label{equ:22}
\end{equation}
where the relevant parameters are selected as: 
$c = 2.997 \times 10^{10}\ \mathrm{cm\ s}^{-1}$, 
$\dot{M}_0 = 2 \times 10^{-6}\ M_{\odot}\ \mathrm{yr}^{-1}$, 
$1\ \mathrm{yr} = 3.156 \times 10^7\ \mathrm{s}$, 
$\sigma_{\mathrm{SB}} = 5.67 \times 10^{-5}\ \mathrm{erg\ s}^{-1}\ \mathrm{cm}^{-2}\ \mathrm{K}^{-4}$, 
$h = 6.625 \times 10^{-27}\ \mathrm{ergs}$, 
$k_B = 1.38 \times 10^{-16}\ \mathrm{erg\ K}^{-1}$, 
$M_{\odot} = 1.989 \times 10^{33}\ \mathrm{g}$, 
and the mass of Black Hole $M = 2 \times 10^6\ M_{\odot}$~\cite{He:2022lrc}.
The flux distribution of the accretion disk is plotted according to the above formula, as shown in Figs. \ref{fig:5} and \ref{fig:6}.
From top to bottom, as the observation angle increases, the intensity distribution of the accretion disk becomes asymmetric, which is due to the Doppler effect. For Figs. \ref{fig:5} and \ref{fig:3}, with the increase of the parameter \( l \) in each row, the changes in the \( r \)-constant orbital imaging and the flux distribution of the accretion disk are not obvious. However, in Fig. \ref{fig:4}, with the increase of the parameter \( a \), the constant-r orbital imaging expands outward significantly, and correspondingly, the radiation flux in Fig. \ref{fig:6} becomes larger and larger.

\section{ Quasinormal Modes }
\label{sec:III}

Quasinormal modes are unique damped oscillatory modes exhibited by perturbed fields when gravitational celestial bodies are disturbed. This section adopts the geometric unit system where \(G = M = c = 1\), the multipole quantum number $ l_*=2 $ (distinguished from the regularization parameter $l$). For scalar field perturbations, they can be described by the Klein-Gordon equation.

\begin{equation}
    \frac{1}{\sqrt{-g}} \partial_{\mu} (\sqrt{-g} g^{\mu\nu} \partial_{\nu} \Psi) = 0
    \label{equ:23}
\end{equation}
The covariant form of the electromagnetic field's equation of motion is given by:
 
\begin{equation}
\frac{1}{\sqrt{-g}} \partial_{\nu} \left( F_{\rho \sigma} g^{\rho \mu} g^{\sigma \nu} \sqrt{-g} \right) = 0
\label{equ:24}
\end{equation}
In a spherically symmetric spacetime, the perturbed field $\Psi$ can be expressed by separating the angular and radial parts as:
\begin{equation}
\Psi(t, r, \theta, \phi) = Y(\theta, \phi) \frac{\psi(t, r)}{r}
\label{equ:25}
\end{equation}
where $Y(\theta, \phi)$ is a spherical harmonic function, describing the angular distribution of the perturbed field, and $\frac{\psi(t, r)}{r}$ describes the radial evolution of the perturbed field. By introducing the tortoise coordinate $\mathrm{d}r_* = \frac{\mathrm{d}r}{f(r)}$, the wave equation of the perturbed field can be written as \cite{Konoplya:2011qq}:

\begin{equation}
\frac{\partial^2 \psi(t, r_*)}{\partial t^2} + V(r_*) \psi(t, r_*) = \frac{\partial^2 \psi(t, r_*)}{\partial r_*^2}
\label{equ:26}
\end{equation}
For a Hayward black hole surrounded by a string cloud, when perturbed by a scalar field, its effective potential is expressed as:

\begin{multline}
V(r_*)=V(r(r_*)) = \left(1 - a - \frac{2 M r^2}{2 l^2 M + r^3}\right)\\ \left[ \frac{l_*(l_* + 1)}{r^2} + \frac{\frac{6 M r^4}{(2 l^2 M + r^3)^2} - \frac{4 M r}{2 l^2 M + r^3}}{r} \right]
\label{equ:27}
\end{multline}
The effective potential during electromagnetic - field perturbations is given by:

\begin{equation}
V(r_*)=V(r(r_*))=\left(1 - a - \frac{2 M r^2}{2 l^2 M + r^3}\right)\left(\frac{l_*(l_* + 1)}{r^2}\right)
\label{equ:28}
\end{equation}

Figs. \ref{fig:7} and \ref{fig:8} show, the effective potential exhibits a single-peaked structure for both scalar and electromagnetic perturbations. As the regularization parameter $l$ increases, the peak value of the effective potential rises gradually, indicating that the potential barrier outside the black hole becomes stronger, thereby affecting the propagation and reflection of the perturbation fields. On the other hand, an increase in the string cloud density parameter a rapidly lowers the potential barrier, showing that the string cloud weakens the potential barrier outside the black hole.

Through the analysis of the effective potential, we can intuitively understand the propagation characteristics of the perturbation field outside the black hole. However, the potential barrier alone cannot accurately describe the 
time-dependent decay behavior and characteristic frequencies of the perturbations. To further reveal the dynamical properties of the black hole under external disturbances, it is necessary to employ a numerical evolution method to integrate the perturbation equation in the time domain and obtain the waveform of the perturbation as it evolves over time.

Perform the separation of variables for $\Psi(t,r_*)$:
\begin{equation}
\psi(t, r_*) = e^{-\mathrm{i}\omega t} \varphi(r_*)
\label{equ:29}
\end{equation}
Substituting the above expression into Eq. \ref{equ:22} yields:

\begin{equation}
\frac{d^2 \psi(r_*)}{dr_*^2} + \left[ \omega^2 - V(r_*) \right] \psi(r_*) = 0
\label{equ:30}
\end{equation}
The boundary conditions for Eq. \ref{equ:26}  are:  as \( r_* \to +\infty \), it is a pure outgoing wave \(\psi(x) \sim e^{-i\omega r_*}\), and as \( r_* \to -\infty \) , it is a pure incoming wave \(\psi(r_*) \sim e^{+i\omega r_*}\)~\cite{Schutz:1985km}.
It is difficult to solve Eq. \ref{equ:26} analytically, and semi-analytical or numerical methods are commonly employed.We use the semiclassical WKB method to calculate the quasinormal mode frequencies, employ the finite difference method to compute the dynamic evolution of the perturbed field over time, and then extract the quasinormal mode frequencies via the Prony method.

The core idea of the WKB method is to perform a Taylor expansion of the potential function at its maximum value \( V_0 \), and match its solution with the WKB asymptotic solutions at infinity and at the horizon. The expression of quasinormal modes in the sixth-order WKB approximation~\cite{Konoplya:2003ii,Konoplya:2019hlu}:

\begin{equation}
\frac{i(\omega^2 - V_0)}{\sqrt{-2V_0''}} - \sum_{i = 2}^{6} \Lambda_i = n + \frac{1}{2}, \ (n = 0, \, 1, \, 2 \dots) 
\label{equ:31}
\end{equation}

The basic idea of the finite difference method is to first divide the domain of the problem into a grid, discretize the original problem into a difference scheme, and then convert it into a system of algebraic equations for solution~\cite{Gundlach:1993tp}. Eq. \ref{equ:26} is rewritten as:
\begin{equation}
\frac{\partial^2 \psi(u, v)}{\partial u \partial v} + \frac{1}{4} V(u, v) \psi(u, v) = 0
\label{equ:32}
\end{equation}
Where $ u := t - r_*, \ v := t + r_* $ are light-like coordinates.The above equation can be discretized as:

\begin{equation}
\psi_N = \psi_W + \psi_E - \psi_S - \Delta u \Delta v V(r) \frac{\psi_W + \psi_E}{8}.
\label{equ:33}
\end{equation}
Given \( \psi_W \), \( \psi_E \), and \( \psi_S \), \( \psi_N \) can be solved. $\Delta$ is the step size of coordinates $u$ and $v$. To solve this equation, it is necessary to establish a $(u, v)$ grid matrix in the plane, and then specify the initial conditions and boundary conditions. Here, a Gaussian pulse is chosen as the boundary condition for the initial perturbation:

\begin{equation}
\psi(\mu = u_0, \, v) = A \exp \left[ - \frac{(v - v_0)^2}{\sigma^2} \right]
\label{equ:34}
\end{equation}

\begin{table*}[t]
\centering
\caption{The QNMs frequencies of black holes under scalar field and electromagnetic field perturbations with different parameters.}
\label{tab:2}
\begin{tabular}{rcccc}
\hline
 & \multicolumn{2}{c}{Scalar} & \multicolumn{2}{c}{Electromagnetic} \\
\hline
$(a=0.1)/l$ & WKB & Prony & WKB & Prony \\
\hline
0.3 & 0.414028-0.0775061i & 0.414422-0.0772544i & 0.394133-0.0762336i & 0.394473-0.0760086i \\
0.6 & 0.420896-0.0744811i & 0.421274-0.0742265i & 0.401486-0.0732019i & 0.401799-0.0729692i \\
0.9 & 0.434286-0.0662172i & 0.434667-0.0659696i & 0.416109-0.0645739i & 0.416392-0.0643202i \\
\hline
$(l=0.5)/a$ & WKB & Prony & WKB & Prony \\
\hline
0 & 0.498616-0.0896236i  & 0.499237-0.0891922i  &  0.473794-0.0878316i &       0.474297-0.0874341i \\     
0.1  & 0.420896-0.0744811i  & 0.421274-0.0742265i & 0.401486-0.0732019i      & 0.401803-0.0729677i  \\      
0.2     & 0.34946-0.05986i & 0.349667-0.0597236i & 0.334829-0.0589651i      & 0.33501-0.0588339i \\
\hline
\end{tabular}
\end{table*}

where $A$ is the amplitude of the pulse, $v_0$ is the central position parameter of the Gaussian pulse, and $\sigma$ is the standard deviation. The finite difference method only gives the dynamic evolution of the perturbed field over time, while other methods are required to extract the quasinormal mode frequencies. The Prony method can be used to analyze and identify signals in exponentially decaying systems~\cite{Berti:2007dg,Lutfuoglu:2025hwh}. At a specific position $r_*$, the wave function can be expanded as:

\begin{equation}
\psi(t) \simeq \sum_{i = 1}^{p} C_i e^{-i \omega_i t}
\label{equ:35}
\end{equation}

Figs \ref{fig:9} and \ref{fig:10} illustrate the time-domain profiles of black holes under scalar and electromagnetic field perturbations for different parameters. Similar to most of the properties discussed above, the variation of the parameter $l$ has little effect on the profiles, while the string cloud parameter $a$ significantly alters their behavior. However, regardless of whether $l$ or $a$ increases, the decay rate of the wave function becomes slower with time. To quantitatively study the quasinormal mode frequencies, Table \ref{tab:2} presents the results calculated by the WKB and Prony methods.The real part of the quasinormal modes describes the “speed” of the black hole’s oscillation after being perturbed. As shown in Table \ref{tab:2}, the real part increases with the increase of the parameter $l$, while it decreases with the increase of the parameter $a$. The imaginary part of the quasinormal modes characterizes the damping process during which the black hole returns to its stable state. A negative imaginary part indicates that the black hole is stable. From the table, it can be seen that as either $l$ or $a$ increases, the absolute value of the imaginary part decreases, which corresponds to the fact that in the time-domain profiles, a larger value of $l$ or $a$ leads to a longer relaxation time for the black hole.

\section{Conclusions}
\label{sec:IV}

In this work, we systematically investigated the Hayward black hole surrounded by a cloud of strings, focusing on the effects of the regularization parameter $l$ and the string cloud parameter $a$ on the spacetime geometry, accretion disk imaging, and QNMs. 

Regarding the spacetime geometry and horizon structure. Increasing $l$ enlarges the inner horizon $r_-$ ,while slightly reducing the outer horizon $r_{+}$, reflecting the enhanced de Sitter effect near the core that suppresses the formation of singularities. Meanwhile, photon-sphere radius $r_{\text{ph}}$, critical impact parameter $b_{c}$, and innermost stable circular orbit $r_{\text{isco}}$ all decrease gradually. In contrast, increasing $a$ slightly decreases the $r_-$ , significantly expands the $r_{+}$,  meanwhile, $r_{+}$, $r_{\text{ph}}$, $b_{c}$, and $r_{\text{isco}}$ all increase rapidly. By comparing the shadow radius with the $3\sigma$ confidence interval of M87* \-($2.546M < R_s < 7.846M$), we identified the parameter regions consistent with the observational constraints. When \(a \geq 0.3\), the black hole shadow radius (corresponding to the critical impact parameter \(b_c\)) exceeds the observational upper limit; therefore, subsequent analyses focus on the cases \(a = 0, 0.1, 0.2\).

We employed the Novikov--Thorne model to simulate the thin-disk images and radiation flux distributions. The results show that variations in $l$ have only a minor effect on the disk images, whereas increasing $a$ causes the constant-$r$ orbits of the accretion disk to expand outward and enlarges the overall radiation flux range.
In the analysis of quasinormal modes, both the scalar and electromagnetic perturbation potentials exhibit a single-peak structure, with the scalar potential slightly higher. Using the WKB and time-domain methods, we find that larger $l$ increases the real part of the QNM frequency, enhances the potential barrier, and suppresses the propagation of perturbations, whereas larger $a$ produces the opposite trend. For the imaginary part, both parameters reduce its absolute value, implying that the damping rate becomes slower.

In summary, this study elucidates how the string cloud and regularization parameter jointly shape the observable features of the Hayward black hole, enriching the understanding of regular black holes under external fields and offering theoretical insight for interpreting black hole shadows, accretion emissions, and quasinormal modes in observations.

\section{Acknowledgements}
This research was  supported by the National Natural Science Foundation of China (Grant No.
12265007).
% \bibliography{basename of .bib file}
%\bibliography{apssamp}

\bibliographystyle{apsrev4-2}

\end{document}